\documentclass[journal,10pt,twoside,twocolumn,letterpaper]{IEEEtran}
\ifCLASSOPTIONcompsoc
    \usepackage[caption=false, font=normalsize, labelfont=sf, textfont=sf]{subfig}
\else
\usepackage[caption=false, font=footnotesize]{subfig}
\fi
\usepackage{graphicx, epstopdf}
\usepackage{lipsum}
\usepackage{amssymb}
\usepackage{amsmath}
\usepackage{amsthm}
\usepackage{mathrsfs}
\usepackage{cite}
\usepackage{color}
\usepackage[acronym]{glossaries}
\usepackage[T1]{fontenc}
\usepackage[latin9]{inputenc}
\usepackage{array}
\usepackage{textcomp}
\usepackage{multirow}
\usepackage{flushend}

\IEEEoverridecommandlockouts

\usepackage{tikz}
\usepackage{textcomp}
\usepackage{hyperref}
\usepackage{lipsum}

\newcommand\copyrighttext{%
  \footnotesize Copyright \textcopyright 2015 IEEE. Personal use of this material is permitted. Permission from IEEE must be obtained for all other uses, in any current or future media, including reprinting/republishing this material for advertising or promotional purposes, creating new collective works, for resale or redistribution to servers or lists, or reuse of any copyrighted component of this work in other works
  DOI: {<10.1109/TVT.2018.2848963>}}
\newcommand\copyrightnotice{%
\begin{tikzpicture}[remember picture,overlay]
\node[anchor=south,yshift=10pt] at (current page.south) {\fbox{\parbox{\dimexpr\textwidth-\fboxsep-\fboxrule\relax}{\copyrighttext}}};
\end{tikzpicture}%
}
    
\makeatletter
\newcommand*\titleheader[1]{\gdef\@titleheader{#1}}
\AtBeginDocument{%
  \let\st@red@title\@title
  \def\@title{%
    \bgroup\normalfont\small\centering\@titleheader\par\egroup
    \vskip0.5em\st@red@title}
}

\title{Resource Allocation in SWIPT Networks under a Non-Linear Energy Harvesting Model: Power Efficiency, User Fairness, and Channel Non-Reciprocity}
\author{Ha-Vu Tran$^{1}$,
Georges Kaddoum$^{1}$,
and Kien T. Truong$^{2}$

\thanks{
$^1$Ha-Vu Tran and Georges Kaddoum are with LACIME Laboratory, University of
Qu\'{e}bec, \'{E}TS engineering school, Montreal, Canada. 
Email: \{ha-vu.tran.1@ens.etsmtl.ca, georges.kaddoum@etsmtl.ca.\}

$^2$Kien T. Truong is with Wireless Systems and Applications Laboratory, Posts and Telecommunications Institute of Technologies, Hanoi, Vietnam. Email: \{kientt@ptit.edu.vn\}

}
}

\begin{document}
\makeatother
\titleheader{This is the authors'version of the paper that has been accepted for publication in IEEE Transactions on Vehicular Technology.}
 \maketitle
\copyrightnotice
\begin{abstract}
This paper considers a multi-user simultaneous wireless information and power transfer (SWIPT) system with a non-linear energy harvesting model, in which a multi-antenna base station (BS) estimates the downlink channel state information (CSI) via uplink pilots. Each single-antenna user is equipped with a power splitter. Three crucial issues on resource management for this system include: (i) power-efficient improvement, (ii) user-fairness guarantee, and (iii) non-ideal channel reciprocity effect mitigation. Potentially, a resource allocation scheme to address jointly such issues can be devised by using the framework of multi-objective optimization. However, the resulting problem might be complex to solve since the three issues hold different characteristics. Therefore, we propose a novel method to design the resource allocation scheme. In particular, the principle of our method relies on structuralizing mathematically the issues into a cross-layer multi-level optimization problem. On this basis, we then devise solving algorithms and closed-form solutions. Moreover, to instantly adapt the CSI changes in practice while reducing computational burdens, we propose a closed-form suboptimal solution to tackle the problem. Finally, we provide numerical results to show the achievable performance gains using the optimal and suboptimal solutions, and then validate the proposed resource allocation scheme.
\end{abstract}


\section{Introduction}\label{sec:1-Introduction}
The explosive progress of information and communication technologies (ICT), such as the fifth generation (5G) networks, has resulted in a tremendous demand for energy to prolong the lifetime of devices in wireless networks \cite{Gupta2015, Ekram2015, StefanoBuzzi2016}. Energy harvesting (EH) techniques can be a promising solution, however, one of the main drawbacks of conventional EH networks is the dependence on unstable energy resources, such as solar and wind energy. To overcome this issue, a radio frequency (RF) wireless power transfer technique has been proposed due to the fact that the RF signals having a frequency range from 3kHz to 300 GHz can be used to carry energy \cite{Grover2010, Lu2015, DusitNiyato}. In RF wireless power transfer (WPT) networks, a transmitter can proactively convey RF signals to recharge energy-hungry devices whenever necessary. Especially, by integrating RF wireless power transfer techniques into traditional wireless communications, the research community has witnessed the prompt development of simultaneous wireless information and power transfer (SWIPT) technique \cite{Lu2015, StefanoBuzzi2016,Ekram2015, DusitNiyato}. Over past few years, the SWIPT technique has showed its bright potential of prolonging the lifetime of wireless devices in many applications, such as cellular, wireless sensor, cognitive radio and Internet-of-Things networks \cite{Lu2015, HaVuTranPotentials, SWIPTKaddoum,TuanAnhLe, PhamVietTuan}. 

In SWIPT networks, the performance of RF energy transfer drastically suffers from path loss. In this concern, resource allocation, whose the principle relies on taking advantages of users' diversity to improve the system performance under limited availabe resources, is one of the key solutions. In \cite{Zhang2013, QShi2014}, the authors have addressed beamforming optimization for SWIPT systems. Recently, researchers have developed a non-linear EH model for which analytical EH results tightly match the measured ones in practical systems \cite{Boshkovska2015a, Boshkovska2017}. More specifically, \cite{Boshkovska2017,RuihongJiang2017} have shown a scheme of joint time and power allocation and a scheme of joint beamforming optimization and power splitting for SWIPT networks, respectively, taking a non-linear EH model into account. In \cite{KeXiong2017}, the authors have investigated rate-energy trade-off behaviour for MIMO SWIPT networks under a non-linear EH model.

In recent years, several articles \cite{HJu2014, Chingoska2016, PDD17,PDD171,PhamVietTuan} have exploited the property of channel reciprocity to develop resource allocation schemes for SWIPT and wireless powered communication  (WPC) networks. According to channel reciprocity, reversed time division duplex (TDD)-based systems can benefit from the fact that the downlink CSI can be achieved by estimating the uplink channels. One of the main advantages of applying TDD to SWIPT systems is that the energy consumption of channel estimation at user receivers is reduced and then the user lifetime can be prolonged consequently. In practice, the antennas at the transmitter and receiver sides have distinct RF chains, and need to be calibrated. However, imperfectly calibrating these antennas can impair the channel reciprocity. Thus, designing robust beamformers taking this issue into account is one of the primary solutions. To the best of our knowledge, there has been little prior work on the impact of non-ideal channel reciprocity for SWIPT systems. 

Furthermore, in conventional communication systems, it is well-known that the users with the worse channel quality consumes more energy for uplink transmissions than the ones with the better channel condition.  Indeed, this phenomenon becomes more prominent in WPC networks where distant users who are often associated with worse channel quality not only have less chance to harvest suffcient energy but also spend more energy for uplink transmission than near users do. Some earlier papers \cite{HJu2014, Liu2014,Tabassum2015,Mishra2017} have studied this phenomenon under the doubly near-far effect point of view. More specifically, most of them have dealt with the unfairness between distant and near users in WPCNs by jointly managing power transfer in downlinks and information transmission in uplinks. Particularly, in those works, the user fairness is improved following channel gains. Nevertheless, this might result in inflexible fairness controls for network operators. The issue of user fairness in WPCNs has inspired us to rethink the resource allocation in SWIPT networks. So far, to the best of our knowledge, most of the previous works \cite{TuanAnhLe, PhamVietTuan, Zhang2013, QShi2014, Boshkovska2017, RuihongJiang2017,KeXiong2017} only address the management of downlink SWIPT to meet ID and EH targets without further prediction of energy demands. In practical SWIPT networks however, after receiving information and harvesting energy in downlinks, the users might need to send pilot and information-bearing signals to the BS via uplinks. Similar to WPCNs, this raises an issue of user unfairness in SWIPT networks, referring that distant users spend more energy to connect to the BS than near users do. Therefore, this motivates us to take into account such an issue when designing resource allocation schemes for downlink SWIPT networks. In fact, considering a practical SWIPT network with a non-linear EH model, flexibly guaranteeing the user fairness between users located at different communication ranges while minimizing transmit power and taking the non-ideal channel reciprocity into account is a challenge.

In this paper, we address the above issues for a SWIPT system under a non-linear EH model in which a multi-antenna BS and multiple single-antenna users are considered. Each user employs a power splitter to coordinate the processes of information decoding (ID) and EH. In this model, downlink CSI is estimated at the BS side via an uplink TDD scheme to exploit the channel reciprocity. Moreover, the channels are assumed to be flat fading. Specifically, this work focuses on designing a novel resource allocation scheme able to jointly {\it (i)} save energy by minimizing transmit power, {\it (ii)} account for the user fairness by maximizing the weighted sum of the coverage probability of EH (under the calibration error), and {\it (iii)} mitigate the non-ideal channel reciprocity by minimizing the effect of calibration error on ID performance. Recently, multi-objective optimization has been promoted as an efficient approach to handle resource allocation in SWIPT networks \cite{EmilBjornson2014,Marler2004,DerrickNg2016,YanSun2016}. However, since the considered objectives (i.e. {\it (i)}, {\it (ii)} and {\it (iii)}) have different characteristics, the algorithms in previous works \cite{DerrickNg2016,YanSun2016} based on the weighted Tchebycheff method \cite{Marler2004} might not be adopted for the considered SWIPT system.

In this work, we propose a novel method in which such objectives are structured on a cross-layer multi-level optimization problem. 
In our method, objectives {\it (i)} and {\it (iii)} can be addressed by a physical-layer optimization problem with a multi-level structure. More specifically, objectives {\it (i)} and {\it (iii)} are formulated into a first-level and a second-level problems, respectively.
Besides, objective {\it (ii)} is transformed into a set of constraints of the physical-layer problem whose threshold values are designed by an upper-layer optimization problem. 
The overall problem formulation would be thoroughly presented in section III. On this basis, we reformulate the upper-layer problem into a quadratic optimization problem, and then solve it using a closed-form optimal solution. After that, considering the multi-level problem, we develop a manner to relax the second-level problem into a set of constraints of the first-level one. Accordingly, the resulting multi-level problem is convex and can be conveniently tackled by solvers \cite{Gra2009}. In practice, the CSI changes with time, due to user mobility, yielding the need of updating and resolving the problem. Thus, to maintain a stable system performance while reducing computational burdens at the BS, we further propose a closed-form suboptimal solution to tackle the problem.
Our main contributions can be stated as follows
\begin{itemize}
\item Proposing a novel resource allocation scheme for SWIPT networks based on structuralizing multiple objectives with different characteristics into the cross-layer multi-level optimization approach.

\item Deriving the closed-form approximate expressions of the average SINR, and the coverage probability of EH under the effect of the non-ideal channel reciprocity.

\item Developing a method to solve the cross-layer multi-level optimization problem.

\item Providing a closed-form suboptimal solution with low complexity to deal with time-varying channels. 
\end{itemize}

The remainder of this paper is organized as follows: In Section II, the system model is described. The overall idea of the proposed resource allocation scheme is discussed in Section III. In Section IV, optimal and suboptimal solutions of the cross-layer multi-level optimization problem are presented. In Section V numerical results and discussions are provided. Finally, concluding remarks are put forward in Section VI.

\textit{Notation:} The notation $\mathbb{R}_{+}^{m}$ and $\mathbb{C}^{m\times n}$ represent the sets of $m$-dimensional nonnegative real vector and $m \times n$ complex matrix, respectively. The boldface lowercase $\mathbf{a}$ and uppercase $\mathbf{A}$ denote vectors and matrices, respectively. The superscripts ${\mathbf A}^{T}$ and ${\mathbf A}^{H}$ indicate the transpose and transpose conjugate, respectively. Moreover, symbols $\left|.\right|$, and $\left\Vert .\right\Vert $ stand for the absolute value, vector Euclidean norm, respectively.

\section{System Model}\label{sec:SystemModel}
\begin{figure}[tb]
\centerline{\includegraphics[width=0.5\textwidth]{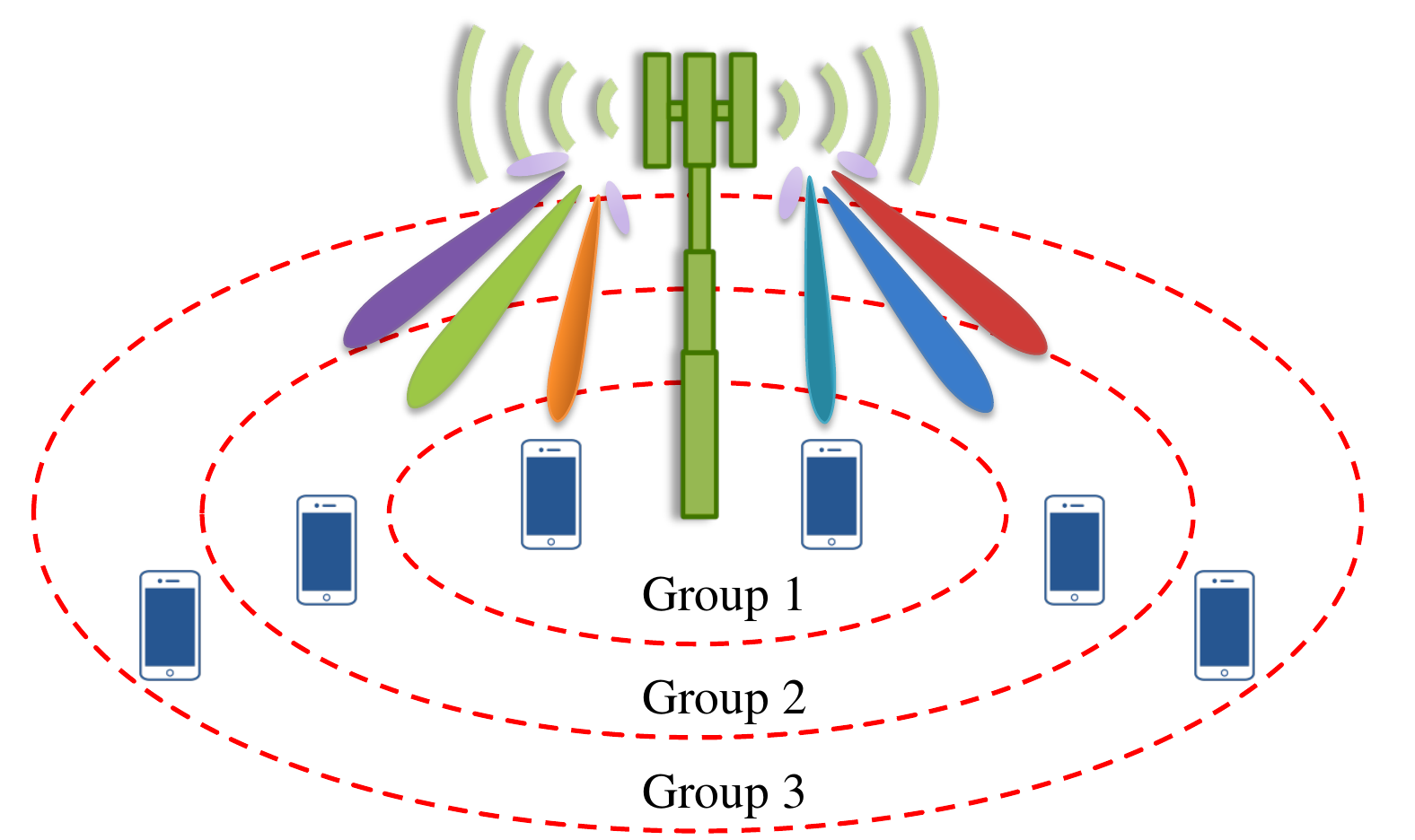}}
\caption{
      An illustration of the SWIPT system including 3 groups of users.
    }
    \label{fig:SystemModel}
\end{figure}
Consider a multiuser SWIPT system when an $M$-antenna BS serves $N$ single-antenna users. The users are randomly located in the cell area such that some users have more severe channel conditions than the other. For example, the users located far from the BS experience higher distance-dependent path-losses than those located near the BS. This might lead to a form of the doubly near-far problem in which distant users not only have less chance to harvest energy but also spend more energy for uplink transmission than near users do. In fact, the effort to improve the EH performance for distant users can lead to some diffculties in managing the ID performance of the overall system since near users might be exposed to denser interference than distant users. Thus, resource allocation is the key solution to this issue.

Therefore, to take into account the doubly near-far problem in providing fairness among users in the cell, we divide the users into $G$ disjoint groups, based on communication ranges, as illustrated in Fig. \ref{fig:SystemModel}. Here, group $g$, $g = 1, \cdots, G$, has $N_g$ users is assigned a unique priority parameter $b_g$. The advantage of prioritizing users can be described as follows. In the previous works of \cite{HJu2014, Liu2014,Tabassum2015}, the EH performance of a user is improved according to channel gains. In practice, however, the channel gain approach may not meet the energy demands exactly, it may offer less or more energy than required. For instance, farthest users whose batteries run out faster may need to harvest more energy than the amount estimated by the channel-gain-based approach. Additionally, when a user is exposed to RF radiation over safety limits, this issue might be diffcult to manage using the channel-gain-based approach. In the approach proposed in this work, the network operator can take other concerns into account and then  flexibly manage the priority to balance system performance.

\subsection{Channel Model of Non-ideal Reciprocity}
We assume that the channels are flat-fading. Let ${\mathbf h}_{n}^g \in {\mathbb C}^{M \times 1}$ and ${\mathbf u}_{n}^g \in {\mathbb C}^{M \times 1}$  be the downlink channel vector and the uplink channel vector between the BS and user $n$, $n=1, \cdots, N_g$, in the group $g$, respectively. According to \cite{3GPP091794}, the downlink and uplink channels can be respectively modelled as 
\begin{align}
{\mathbf h}_{m,n}^g = t_{m} v_{m,n}^g {\bar r}_{n}^g,\\
{\mathbf u}_{m,n}^g = r_{m} v_{m,n}^g {\bar t}_{n}^g,
\end{align}
where $v_{n,m}^g$, $m = 1, \cdots, M$, is the wireless channel coefficient between transmit antenna $m$ of the BS and user $n$ in group $g$; $t_{m}$ and $r_{m}$ are the equivalent transmit and receive antenna gains of antenna $m$ of the BS, respectively; ${\bar t}_{n}^g$ and ${\bar r}_{n}^g$ are the equivalent receive and transmit circuit gains of user $n$ in group $g$, respectively.

Let us define $c_{m,n}^g$ such that $\dfrac{{\mathbf h}_{m,n}^g}{{\mathbf u}_{m,n}^g} = \dfrac{t_{m} {\bar r}_{n}^g}{r_{m} {\bar t}_{n}^g} = c_{m,n}^g$. If antennas are perfectly calibrated, then ${c}^{g}_{1,n} = {c}^{g}_{2,n}=\cdots = {c}^{g}_{M,n} = \tilde c^{g}_{n}$, where $\tilde c^{g}_{n}$ is a constant number  \cite{3GPP091794}. In this case, channel reciprocity holds and the BS is able to estimate the downlink channels based on uplink pilots perfectly. In fact, some papers on TDD channel reciprocity observe that the difference between the uplinks and downlinks, i.e. $ \tilde c^{g}_{n}$, does not affect the designs of beamforming/precoding/detector in multi-user multi-input multi-output (MIMO) systems \cite{Vieira14}.

However, there are always hardware calibration errors in practice. Thus, according to one of the most common approaches \cite{3GPP091794}, the downlink channels can be modelled considering calibration errors as follows 
\begin{align}\label{eq:CEE}
\mathbf{ h}^{g}_{n} = (\mathbf{I}_{M} + \mathbf{C}^{g}_{n})\mathbf{u}^{g}_{n},
\end{align}
where $\mathbf{I}_{M}$ is an $M \times M$ identity matrix, and $\mathbf{C}^{g}_{n} = \text{diag}\{c^{g}_{1,n}, c^{g}_{2,n},\cdots, c^{g}_{M,n}\}$ is a matrix presenting calibration errors and $c^{g}_{m,n} \sim \mathcal{CN}(0, \sigma^{2}_{cal})$.

In practice, users may have more than one antenna. Thus, we provide a brief discussion on the extension of channel model as follows. Since both the BS and users are equipped multiple antennas, the hardware calibration might be complex. In light of \cite{3GPP091794}, to reduce the compexity, it is suggested calibration be mainly performed at the BS side while the antennas at the user side are pre-calibrated such that $ \dfrac{{\bar r}_{n,l}^g}{{\bar t}_{n,l}^g} = {\lambda_{n}^g}$ $(\forall l)$ where the index $l$ indicates antenna $l$ at a user and $\lambda_{n}^g$ is a constant. Hence, we obtain an expression $\dfrac{{\mathbf h}_{m,n,l}^g}{{\mathbf u}_{m,n,l}^g} = \dfrac{t_{m} {\bar r}_{n,l}^g}{r_{m} {\bar t}_{n,l}^g} = {c_{m,n}^g}$. Note that index $l$ does not appear in ${c_{m,n}^g}$ since antenna $l$ has been pre-calibrated. Then, similar to the case of single-antenna user, the channel model can be formulated as
\begin{align}
[ \mathbf{ h}^{g}_{n,1} \quad  ... \quad \mathbf{ h}^{g}_{n,L} ] = (\mathbf{I}_{M} + \mathbf{C}^{g}_{n})[
   \mathbf{ u}^{g}_{n,1} \quad ...  \quad \mathbf{ u}^{g}_{n,L} ].
\end{align}

\subsection{Signal Model}
On the downlink, the received signal at user $n$ in group $g$ could be expressed as
\begin{align}\label{eq:RSig}
y_n^g &= ({{\mathbf h}_{n}^g})^T {\mathbf w}_n^g  s_n^g + \sum\limits _{{n'=1} \atop ({n'} \ne n)}^{N_g}  ({ {\mathbf h}_{n}^{g} })^T  {\mathbf w}_{n'}^{g}   s_{n'}^{g} \\ \nonumber
&+ \sum\limits _{g'=1 \atop (g' \ne g)}^{G} \sum\limits _{n=1}^{N_g}  ({  {\mathbf h}_{n}^{g} } )^T {\mathbf w}_n^{g'} s_n^{g'}  + n_0,
\end{align}
where ${\mathbf w}_n^g$ is the beamforming vector for user $n$ in group $g$, and $s_n^g$ is the unit power transmit symbol. The first term is the desired signal, the second term is the interference in the same group, the third term is the interference from the other groups, and the fourth term $ n_0$ is the additive white Gaussian noise (AWGN), i.e. $n_0 \sim \mathcal {CN} (0, \sigma^2_{0})$. 

Further, the received signal at user $n$ in group $g$ is split to the information decoder and the energy harvester by a power spitter, which divides an $\rho^g_n$ $(0 \le \rho^g_n \le 1)$ portion of the signal power to the information decoder, and the remaining $(1 - \rho^g_n)$ portion of power to the energy harvester. As a result, the signal split to the information decoder is expressed as
\begin{align}
y_n^{\text{ID},g} = \sqrt{\rho^g_n} y_n^g +  n_1,
\end{align}
where $n_1$ is the AWGN introduced by the ID, i.e. $n_1 \sim \mathcal {CN} (0, \sigma^2_{1})$. Due to calibration errors, the BS has imperfect downlink CSI, and the instantaneous signal-to-interference-plus-noise ratio (SINR) values are not available. Alternatively, to evaluate ID performance, the average SINR at user $n$ in group $g$ can be given by
\begin{align}\label{eq:SINR}
\mathbb E \left[ \mathtt {SINR}^g_n \right] { \buildrel \Delta \over =    \dfrac{ \mathbb E \left[  \rho^g_n \left| ({{\mathbf h}_{n}^g})^T {\mathbf w}_n^g \right|^2 \right]} {\mathbb E \left[ \rho^g_n \left( \mathcal I_n^g + \left| n_0 \right|^2 \right) + \left| n_1 \right|^2  \right]}},
\end{align}
where $\mathcal I_n^g$ is the interference component which can be presented as below
\begin{align}\label{eq:Interference}
\mathcal I_n^g = \sum\limits _{n'=1 \atop (n' \ne n)}^{N_g}  \left| ({ {\mathbf h}_{n}^{g} })^T  {\mathbf w}_{n'}^{g} \right|^2 + \sum\limits _{g'=1 \atop (g' \ne g)}^{G} \sum\limits _{n'=1}^{N_g}  \left| ({  {\mathbf h}_{n}^{g} })^T {\mathbf w}_{n'}^{g'}  \right|^2.
\end{align}
It is worth noting that we apply a relaxation for computing the average SINR given in \eqref{eq:SINR} due to further conveniences.

\subsection{Non-Linear Energy Harvesting Model}
Further, the signal split to the energy harvester is 
\begin{align}
y_n^{\text{EH},g} = \sqrt{(1-\rho^g_n)} y_n^g.
\end{align}

Conventionally, in linear EH models, the harvested energy, denoted by $\mathtt {E}_n^{g}$, can be computed as a linear function of input energy, i.e. $\mathtt {E}_n^{g} = \xi^g_n \mathtt {\hat E}_n^{g} $ in which $\xi^g_n$ $(0 \le \xi^g_n \le 1)$ is the energy conversion efficiency and $\mathtt {\hat E}_n^{g}$ is the input energy defined by
\begin{align}\label{eq:PCEH1}
\mathtt {\hat E}_n^{g} =  {(1-\rho^g_n)} \sum\limits _{g'=1 }^{G} \sum\limits _{n'=1}^{N_g}\left| ({  {\mathbf h}_{n}^{g} })^T  {\mathbf w}_{n'}^{g'} \right|^2.
\end{align}
Note that the contribution of $n_0$ and $n_1$ is neglected \cite{Zhang2013,Ng2014}.

However, in practice, it has been shown that the RF energy conversion efficiency varies with different levels of input energy.
The linear model is only proper for the specific scenario when the received powers at users are constant. Thus, in this work, we consider a non-linear EH model that can give more accurate results to practical systems. According to \cite{Boshkovska2017,RuihongJiang2017,KeXiong2017}, the non-linear model can be described as
\begin{align}\label{eq:PCEH2}
\mathtt {E}_n^{g}= \dfrac{ \dfrac{\mathtt{M}^{EH}}{1 + e^{-\mathtt{a}(\mathtt {\hat E}_n^{g} -\mathtt{b})} } - \dfrac{\mathtt{M}^{EH}}{1 + e^\mathtt{ab}} }{1 - \dfrac{1}{1 + e^\mathtt{ab}}},
\end{align}
where $\mathtt{M}^{EH}$ is maximum harvested energy at a user when the EH circuit meets saturation. In addition, $\mathtt{a}$ and $\mathtt{b}$ are constants regarding circuit specifications, e.g. the
resistance, capacitance, and diode turn-on voltage \cite{Boshkovska2017,RuihongJiang2017,KeXiong2017}.

In this work, to characterize the chance of harvesting energy at a user, the coverage probability of harvested energy regarding calibration errors \cite{GrahamUpton} is given by 
\begin{align}\label{eq:PCEH3}
\mathcal Pr \left(\mathtt {E}_n^{g} \ge \theta_n^g \right) ,
\end{align}
where $ \theta_n^g$ is a threshold.

\section{Problem Formulation and Proposed Approach}
This section describes the overall idea of the proposed resource allocation scheme and presents the related problem formulation.

We start with discussing the considered objectives for resource allocation in the SWIPT system.
First, one of the important objectives is transmit power saving, namely objective {\it (i)}. Second, the overall EH performance should be maximized taking the user fairness into account, namely objective {\it (ii)}. Third, the effect of non-ideal channel reciprocity on ID performance should be minimized, namely objective {\it (iii)}. To handle this task, although multi-objective optimization seems to be promising \cite{EmilBjornson2014,Marler2004}, directly applying it might bring non-trivial computing difficulites. For instance, one of the most popular methods, so-called {\it weighted sum method} \cite{Marler2004}, might not be a suitable choice since the characteristic of the objectives are not the same. Indeed, the metrics of objectives $(i)$, $(ii)$, and $(iii)$ are Watt(s), percentage(s), and dB. This leads to difficulties of choosing weighted values \cite{Marler2004}. Further, another multi-objective optimization method, namely {\it lexicographic method} \cite{Marler2004}, can deal with this drawback. However, the resulting problem is highly complex and hard to solve.

In this work, we propose a novel method o jointly solve these objectives. In fact, objective {\it (ii)} can be represented by maximizing the weighted sum of coverage probability of EH according to the group priorities. In this concern, distant users with a higher priority can be managed to have more chance to harvest energy than near users. Thus, our idea exploits this point and then proposes a cross-layer design in terms of a cooperation between the physical and the upper layers. In this concern, objectives {\it (i)} and $(iii)$ are jointly formulated into the physical-layer problem. Besides, objective {\it (ii)} is translated into the EH coverage probability constraints of the physical-layer problem whose threshold values are planned by an upper-layer optimization problem, namely OP$_2$. After solving {OP}$_2$, the designed plan of the thresholds is conveyed to the physical layer. The proposed approach has many advantages, of which we include three. First, the issue of dissimilarity of characteristics between the objectives can be properly relaxed when the objective {\it (ii)} is separately tackled at the upper layer. Second, the computation burden is shared between the processing units of the layers. Third, network operators can  flexibly manage the user fairness and balance the system performance since they can check the designed plan before conveying it to the processing unit at the physical layer. The detailed formulation is shown below.

Accounting for problem OP$_2$, we consider a realistic situation where the network operator has an existing serving plan regarding the thresholds of the EH coverage probability constraints. This plan might be created according to user demands or registered Quality of Service (QoS) only. It does not take the user fairness into account. Thus, re-designing the plan is to address this task. Moreover, in practice, the network operator might not want the new plan to deviate too much from an existing plan. In this concern, the first aim is to maximize the weighted sum of the threshold with the purpose of guaranteeing the user fairness, whereas the second one is to minimize the 0-norm of the difference between the existing plan and the re-designed one. Thus, based on the user-prioritized system model, re-designing the plan can be represented by a multi-objective optimization problem as follows
\begin{subequations}\label{eq:MOProblem0}
\begin{align}
\text{OP}_2 : \underset{\{\alpha^g_n\}} {\max} \quad  &\sum_{g=1}^{G} \sum_{n=1}^{N_g} b^{g} \alpha^g_n \quad \&  \quad 
		      \underset{\{\alpha^g_n\}} {\min} \quad {\left\| \pmb \alpha -  {\mathbf q} \right\|_0 } \label{eq:MOProblem0k}
\\
			  \text{s.t.:}  \quad & \sum_{g=1}^{G} \sum_{n=1}^{N_g} \alpha^g_n = \sum_{g=1}^{G} \sum_{n=1}^{N_g}  q^g_n, \label{eq:MOProblem0m}
\\
			\quad &	0 \le \alpha^g_n \le 1, \quad \forall g,n, \label{eq:MOProblem0j}
\end{align}
\end{subequations}
in which, $\{q^g_n\}$ and $\{\alpha^g_n\}$ represent the threshold values of coverage probabilites set in the existing and desired plans, respectively. Also, ${\pmb \alpha} = [\alpha_1^1 \ldots \alpha_N^G]^T$ and ${\mathbf q} = [q_1^1 \ldots q_N^G]^T$. It is worth mentioning that $b_g$ is the priority factor associated with group $g$. Additionally, constraint \eqref{eq:MOProblem0m} is to keep the total coverage probability (i.e. the total offered QoS of the system) given in the two plans equal. 

\begin{figure}[!]
\centerline{\includegraphics[width=0.37\textwidth]{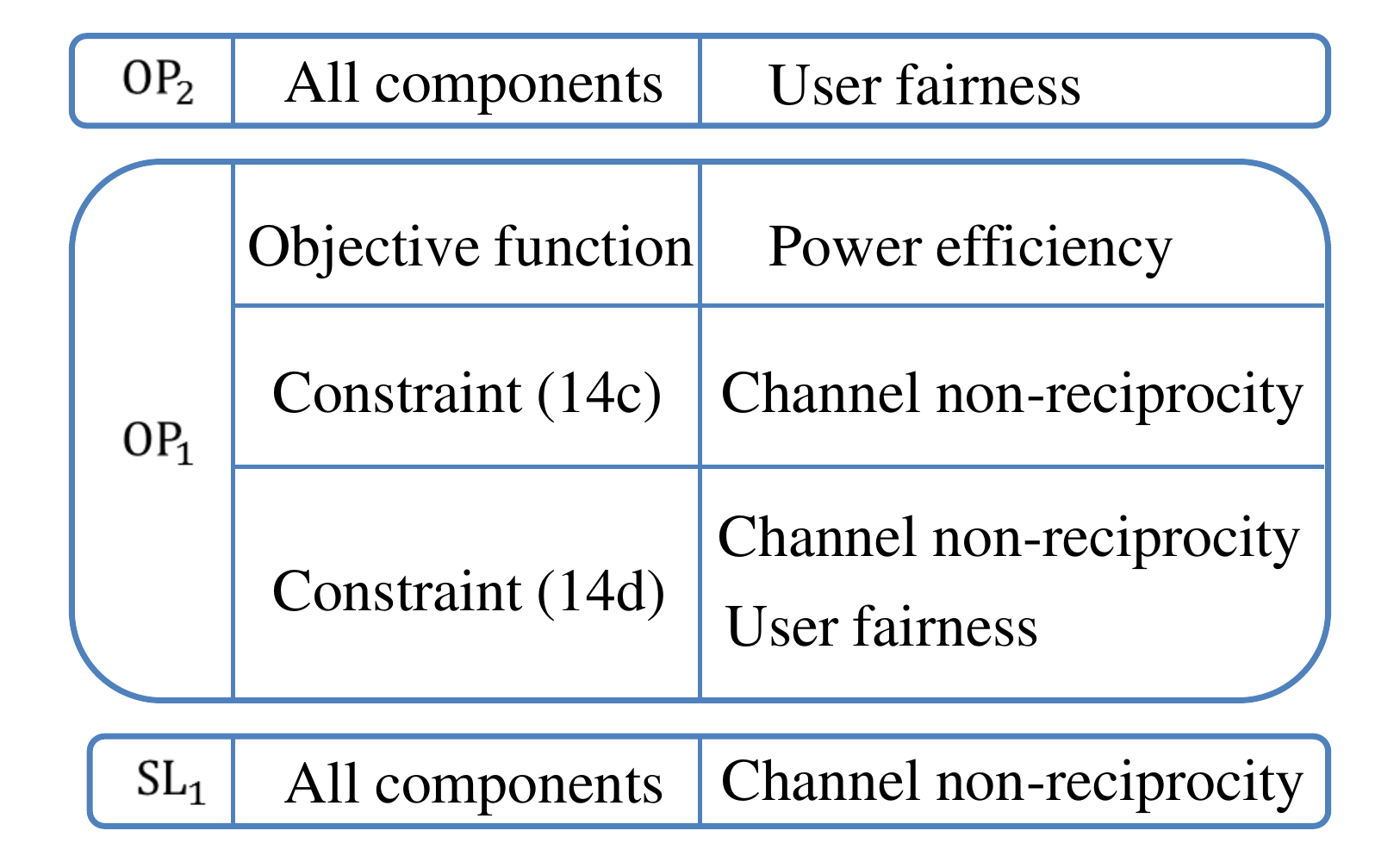}}
\caption{
     {Responsibility for each objective function and each constraint.}
    }
    \label{fig:task}
\end{figure}

Considering the physical-layer problem, on one hand, objective {\it (i)} can be formulated by minimizing total transmit power. On the other hand, to combat the channel reciprocity effect on the ID performance (i.e. objective $(iii)$), it is desirable to design robust beamformers to maximize the average SINR obtained at each user. However, since the beamforming vectors for multiple users in the expression of the average SINR are coupled. This might make this multi-objective problem intractable. Thus, we propose an alternative manner to manage objectives $(i)$ and $(iii)$ using a multi-level structure. More specifically, the first-level problem, namely OP$_1$, is to account for objective {\it (i)}, whereas the second-level problem, namely SL$_1$, is to address objective $(iii)$. However, even in this case, achieving the beamformers such that the beam direction and beam power are jointly optimized is difficult. Thus, considering problem SL$_1$, it is suggested that only the beam directions of beamformers are designed to maximize the average signal-to-leakage ratio (SLR) on calibration errors at each user with setting the average SINR as constraints. In this concern, specifically, SLR beamforming is also known by another name, transmit minimum mean square error (MMSE) beamforming [Remark 3.2, \cite{EBjornson2013}]. Its principle relies on minimizing the effect of calibration errors on the desired signal performance. Besides, it is well-known that the leakage or the interference can be useful in improving EH performance. Thus, the SLR criterion might be non-preferred to EH. In fact, dealing with non-ideal channel reciprocity to improve ID performance, in its nature, affects the EH performance. This can be seen as a performance trade-off between ID and EH. However, benefting from the cross-layer approach, the EH performance can be guaranteed through the thresholds designed by upper-level problem OP$_2$. 

For further concerns, detailed problem formulations can be presented as follows
\begin{subequations}\label{eq:MOProblem}
\begin{align}
	\text{OP$_1$:} \quad  \underset{\{{\mathbf w}^g_n, \rho^g_n \}}\min & \quad   \sum_{g=1}^{G} \sum_{n=1}^{N_g} \left\| {{\mathbf w}^g_n} \right\|^2 \label{eq:MOProblema} 
\\
	\text{s.t.:}   \quad & {\{{\mathbf w}^g_n\}} \in \mathcal F, \label{eq:MOProblemb}			     
\\			  
			  \quad & \mathbb E \left[ \mathtt {SINR}^g_n \right] = \gamma^g_n, \quad \forall g,n, \label{eq:MOProblemf}
\\
			\quad & {\mathcal Pr} \left( \mathtt {E}^g_n \ge \theta^g_n \right)  \ge \alpha^g_n, \quad \forall g,n, \label{eq:MOProblemg}
\\			
			\quad & 0 < \rho^g_n < 1, \quad \forall g,n, \label{eq:MOProblemn}
\end{align}
\end{subequations}
where $\{\gamma^g_n\}$, $\{\theta^g_n\}$, and $\{\alpha^g_n\}$ are the thresholds of the average SINR, the EH, and the coverage probability of EH, respectively. In particular, the values of $\{\alpha^g_n\}$ are designed by the upper layer, i.e. problem OP$_2$, such that the weighted sum of the threshold values is maximized. $\{\alpha^g_n\}$ can be affected by the method used to solve multi-objective problem OP$_2$, determined by network operators. Additionally, the values of $\{\gamma^g_n\}$ and $\{\theta^g_n\}$ can be also assigned by network operators. In practice, the network operators can determine these thresholds based on several concerns, such as user requests and the registered QoS at the user side. Further, $\mathcal F$ is a set defined by the second-level problem SL$_1$.

\begin{figure}[!]
\centerline{\includegraphics[width=0.38\textwidth]{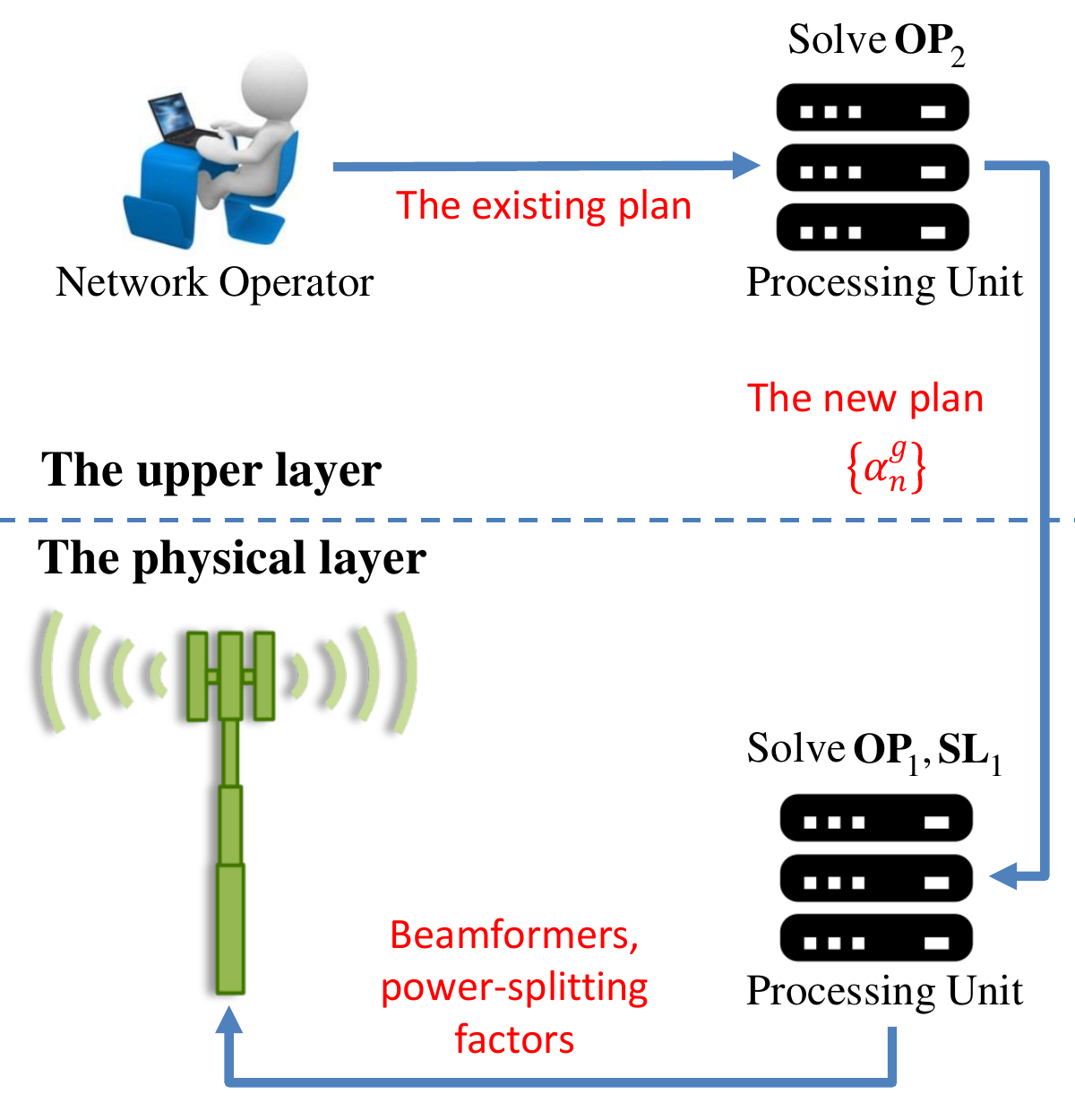}}
\caption{
      The proposed resource allocation scenario.
    }
    \label{fig:flowchart}
\end{figure}

Regarding the formulation of second-level problem SL$_1$, the beamformer should be designed to maximize the average SLR at each user. Accordingly, problem SL$_1$ can be formulated as below
\begin{align}\label{eq:ProblemSL1x}
\text{SL}_1: & \left\{ \underset{{{\mathbf w}^g_n}} {\max} \quad  \overline {\mathtt {SLR}}_n^g \right\}, \quad \forall n,g 
\end{align}
where
\begin{align}\label{eq:SLR}
&\overline {\mathtt {SLR}}_n^g = \\
&\frac{ ({{\mathbf w}^g_n})^H \mathbb E \left[ {{\mathbf h}_{n}^g} ({{\mathbf h}_{n}^g})^H \right] {{\mathbf w}^g_n} } { ({{\mathbf w}^g_n})^H \mathbb E \left[ \sum\limits_{n'=1 \atop (n' \ne n)}^N    {{\mathbf h}_{n'}^g} ({{\mathbf h}_{n'}^g})^H   +  \sum\limits_{g'=1 \atop (g' \ne g)}^G \sum\limits_{n=1}^N  {{\mathbf h}_{n}^{g'}}  ({{\mathbf h}_{n}^{g'}})^H   \right]{{\mathbf w}^g_n}}.
\end{align}
in which, the numerator includes the desired signal power while the denominator consists of the total interference power. In principle of the multi-level problem \cite{Caramia2008}, constraint \eqref{eq:MOProblemb} implies that any optimal solutions $\{{{\mathbf w}^g_n}\}$ of the first-level problem must belong to the set of the optimizers of second-level problem SL$_1$, i.e. $\mathcal F$.  In other words, optimal beamformers, i.e. $\{{\mathbf w}^{\star g}_n\}$, should have the beam directions designed by problem SL$_1$. For convenience, responsibility of each problem and an illustration of the overall proposed scheme are shown in Figs. \ref{fig:task} and \ref{fig:flowchart}, respectively.

\section{Solutions to the Cross-Layer Multi-Level Optimization Problem }\label{sec:Solution}
This section is to present a novel method to tackle the cross-layer multi-level problem. 

\subsection{Solving Upper-Layer Problem OP$_2$}
On one hand, it is obvious that our aim is to design a new serving plan taking the user fairness into account. This is hence formulated into OP$_2$ which is a multi-objective optimization problem.
Due to the aim of the new plan, the first objective, i.e. maximizing the weighted-sum of the coverage probabilities, should be treated more importantly than the second objective, i.e. minimizing the difference between the two plans.

On the other hand, actually, solving a single-objective problem with the $l_0$-norm is difficult. This is even more challenging when considering this issue in the form of a multi-objective problem.  In principle, the $l_0$-norm handles the plan difference through the number of dissimilar elements between the two plans. However, this approach results in a strongly NP-hard problem which is highly complex and computationally intractable to solve \cite{Ge2011, Baraniuk2007}. The main difficulty is due to the sparsity property of the $l_0$-norm. Several previous works proposed methods to relax the $l_0$-norm into a higher-order norm problem \cite{FenChen2013, Iordache2014}. However, it is still difficult to adopt directly those methods to the multi-objective problem.

As discussed, since the second objective is less important than the first one, our idea is to exploit this point to relax the sparsity property required in the second objective and then make problem OP$_2$ more tractable. In this concern, we propose minimizing the $l_2$-norm of the plan difference as an alternative approach. In fact, the $l_2$-norm might not measure the sparsity of the plan difference, however, it can measure the sum of the plan difference which the $l_0$-norm might not do. Particularly, the previous work [\cite{DAgarwal}, pp. 488] managed the plan difference by minimizing the squared variance (i.e. a form of the $l_2$-norm) between the two plans. Thus, we focus on minimizing the squared variance, instead of minimizing the $l_0$-norm of the plan difference. Accordingly, problem OP$_2$ is reformulated as follows

\begin{subequations}\label{eq:MOProblem}
\begin{align}
\text{OP}_2 : \underset{\{\alpha^g_n\}} {\max} \quad  &\sum_{g=1}^{G} \sum_{n=1}^{N_g} b^{g} \alpha^g_n \quad \&  \quad 
		      \underset{\{\alpha^g_n\}} {\min} \quad \sum_{g=1}^{G} \sum_{n=1}^{N_g} (\alpha^g_n - q^g_n)^2  \label{eq:MOProblemk}
\\
			  \text{s.t.:}  \quad & \sum_{g=1}^{G} \sum_{n=1}^{N_g} \alpha^g_n {=} \sum_{g=1}^{G} \sum_{n=1}^{N_g}  q^g_n, \label{eq:MOProblemm}
\\
			\quad &	0 \le \alpha^g_n \le 1, \quad \forall g,n. \label{eq:MOProblemj}
\end{align}
\end{subequations}

Based on suggested methods given in \cite{Marler2004}, multi-objective problem OP$_2$ can be solved by using the weighted-sum approach. In this regard, we introduce a parameter $R$ $(0<R<1)$ to represent the relative importance between the two objectives. Hence, OP$_2$ can be re-written as 
\begin{align}\label{eq:MOProblemOP2}
\text{OP}_{2-1} : \underset{\alpha^g_n} {\max} \quad  & {(1-R)}\sum_{g=1}^{G} \sum_{n=1}^{N_g} b^{g} \alpha^g_n - R \sum_{g=1}^{G} \sum_{n=1}^{N_g} (\alpha^g_n - q^g_n)^2 
\\
			  \text{s.t.:}  \quad & \eqref{eq:MOProblemm}, \eqref{eq:MOProblemj}. \nonumber
\end{align}
By modifying $R$, a trade-off between the objective functions can be evaluated. On this basis, a resource allocation scheme can be reasonably designed, according to network requirements. To derive the closed-form solution, the Lagrangian function is derived as
\begin{align}\label{eq:SL2-2}
&\mathcal L^{s2} \left( {\pmb \alpha}, \mu^{s2}, {\pmb \delta}^{s2}_1, {\pmb \delta}^{s2}_2 \right) \nonumber \\
&= - {\pmb \alpha}^T {\mathbf f} 
+ {R {\pmb \alpha}^T {\pmb \alpha}} + \mu^{s2} \left( {\pmb \alpha}^T {\mathbf 1}_{NG} - \sum_{g=1}^{G} \sum_{n=1}^{N_g}  q^g_n  \right) \nonumber \\
&+ \sum_{g=1}^{G} \sum_{n=1}^{N_g}  (q^g_n)^2 + ({\pmb \delta}^{s2}_1)^T \left( {\pmb \alpha} - {\mathbf 1}_{NG} \right) - ({\pmb \delta}^{s2}_2)^T {\pmb \alpha}.
\end{align}
where ${\mathbf f} = \left[ f_1^1 \ldots f_{N_G}^1  \ldots  f_1^G  \ldots  f_{N_G}^G  \right]^T \in \mathbb C^{NG \times 1}$ in which $f_n^g = (1-R)b^g + 2{Rq_n^g}$, ${\pmb \alpha} = \left[ \alpha_1^1 \ldots \alpha_{N_G}^1  \ldots  \alpha_1^G  \ldots  \alpha_{N_G}^G  \right]^T \in \mathbb C^{NG \times 1}$, $\mu^{s2} \in \mathbb R$, ${\pmb \delta}^{s2}_1 \in \mathbb R^{NG \times 1}$ and ${\pmb \delta}^{s2}_2 \in \mathbb R^{NG \times 1}$ are dual variables. After some manipulations, the optimal solution, i.e. ${\pmb \alpha}^{\star}$, can be obtained by
\begin{align}
{\pmb \alpha}^{\star} &= \frac{1}{2R} \left( {\mathbf  f}  -\mu^{s2}{\mathbf 1} - {\pmb \delta}^{s2}_1 + {\pmb \delta}^{s2}_2 \right). 
\end{align}

By substituting ${\pmb \alpha}^{\star}$ into the Lagrangian function, we can obtain the dual problem. 
To make mathematical presentation favorable, we re-arrange the dual variables into only one vector, and then re-write the above equation in a more convenient form as follows
\begin{align}
{\pmb \alpha}^{\star} = \frac{1}{2R} \left( {\mathbf  f} + {\mathbf  B}{\mathbf  d} \right),
\end{align}
in which ${\mathbf B} = \left[ -{\mathbf  1}_{NG} \quad -{\mathbf  I}_{NG} \quad {\mathbf  I}_{NG}  \right]$ and ${\mathbf  d} = \left[ \mu^{s2} \quad ({\pmb \delta}^{s2}_1)^T \quad ({\pmb \delta}^{s2}_2)^T \right]^T$. Next, we plug ${\pmb \alpha}^{\star}$ back into the Lagrangian function given in \eqref{eq:SL2-2}. After some manipulations, we have
\begin{align}
\mathcal L^{s2} \left( {\pmb \alpha}^{\star}, \mu^{s2}, {\pmb \delta}^{s2}_1, {\pmb \delta}^{s2}_2 \right) 
= {\mathbf  s}^T{\mathbf d} - \frac{1}{4R} \left( {\mathbf f} + {\mathbf B}{\mathbf d} \right)^T \left( {\mathbf  f} + {\mathbf  B}{\mathbf  d} \right),
\end{align}
where ${\mathbf s} = \left[ -\left(\sum_{g=1}^{G} \sum_{n=1}^{N_g}  q^g_n \right)  \quad -{\mathbf  1}_{NG}^T \quad {\mathbf  0}_{NG}^T \right]^T$. Accordingly, the dual problem can be formulated as
\begin{subequations}\label{eq:SL2-3}
\begin{align}
\underset{{\mathbf  d}} \min \quad &  {\mathbf  s}^T{\mathbf d} - \frac{1}{4R} \left( {\mathbf  f} + {\mathbf  B}{\mathbf d} \right)^T \left( {\mathbf  f} + {\mathbf B}{\mathbf  d} \right)
\\
	\text{s.t.}   \quad & {\mathbf  d} \succeq 0.
\end{align}
\end{subequations}
Since problem OP$_{2-1}$ is quadratic convex and all constraints are convex, there never exists a duality
gap \cite{StephenBoyd2004}.

In order to transform \eqref{eq:SL2-3} into a more tractable form, we let $\hat {\mathbf  B} = {\mathbf  B}^T{\mathbf B}/R$ and $\hat {\mathbf  f} = {\mathbf s} - {\mathbf  B}^T{\mathbf  f}/{R}$. On this basis, the above problem can be represented in the well-known form
\begin{subequations}\label{eq:SL2-4} as
\begin{align}
\underset{{\mathbf d}} \min \quad &  \frac{1}{4} {\mathbf d}^T \hat {\mathbf B}  {\mathbf  d} - \hat {\mathbf  f}^T {\mathbf  d} \label{eq:SL2-4a}
\\
	\text{s.t.}   \quad & {\mathbf  d} \succeq  {\mathbf 0} \label{eq:SL2-4b}.
\end{align}
\end{subequations}

Indeed, problem \eqref{eq:SL2-4} is derived in a convex formulation. However, due to nonnegativity constraint \eqref{eq:SL2-4b}, there might not exist analytical solutions of optimal ${\mathbf d}^{\star}$. Hence, an iterative algorithm would be needed. Conventionally, gradient iteration algorithm is applied to update the Lagrangian multipliers. However, the work of \cite{Sha2007} has proven that multiplicative updates can enhance the value of the problem objective at each iteration and hence monotonically converge to the optimal solution. This motivates us to construct an algorithm in terms of multiplicative updates to tackle the problem \eqref{eq:SL2-4}. Then, we define $\hat {\mathbf B}^{ \bullet}$ and $\hat {\mathbf B}^{\circ}$ as follows
\begin{align}\label{eq:SL2-5}
\hat {B}^{\bullet}_{ij} = \left\{ {\begin{array}{*{20}{l}}
\hat {B}_{ij} \quad &\text{if} \quad \hat {B}_{ij} \ge 0,  \\
0  \quad &\text{if} \quad \text{otherwise},
\end{array}} \right.
\end{align}
\begin{align}\label{eq:SL2-6}
\hat { B}^{\circ}_{ij} = \left\{ {\begin{array}{*{20}{l}}
|\hat {B}_{ij}| \quad &\text{if} \quad \hat {B}_{ij} \le 0,  \\
0  \quad &\text{if} \quad \text{otherwise},
\end{array}} \right.
\end{align}

On this basis, the iterations of multiplicative updates can be shown as
\begin{align}\label{eq:SL2-6}
d_i  \leftarrow  d_i \left( \frac{\hat f_i + \sqrt{\hat f_i^2 + ({\hat {\mathbf B}^{\bullet}} {\mathbf d})_i ({\hat {\mathbf B}^{\circ}} {\mathbf d})_i  } }{ ({\hat {\mathbf B}^{\bullet}} {\mathbf d})_i} \right), \quad \forall i.
\end{align}
The convergence of the iterative algorithm is ensured \cite{Sha2007}. Through achieving optimal solution ${\mathbf d}^{\star}$, the optimal values of Lagrangian multipliers, i.e. $\mu^{\star s2}, {\pmb \delta}^{\star s2}_1, {\pmb \delta}^{\star s2}_2$, can be conveniently found. As a result, optimal value ${\pmb \alpha}^{\star}$ is determined.

\subsection{Solving Second-Level Problem SL$_1$}
In fact, the beamforming vector can be decomposed into two components which are the beam power and the beam direction. In the mathematical presentation, this can be shown as
\begin{align}\label{eq:theta2}
{\mathbf w}_n^g = \sqrt{p_n^g} {\pmb \nu}_n^g, \quad \forall g,n,
\end{align}
where $\left\| {\pmb \nu}^g_n \right\|^2 = 1$. Without loss of generality, we can eleminate parameter $p_n^g$ at both the numerator and denominator of the average SLR given in \eqref{eq:SLR}. Accordingly, problem SL$_1$ can be equivalently reformulated as below
\begin{subequations}
\begin{align}\label{eq:ProblemSL1x}
\text{SL}_1: & \left\{ \underset{{\pmb \nu}^g_n} {\max} \quad  \overline {\mathtt {SLR}}_n^g \right\}, \quad \forall n,g 
\\
			 &\quad \text{s.t.:}  \left\| {\pmb \nu}^g_n \right\|^2 = 1, \quad \forall g,n,
\end{align}
\end{subequations}
where
\begin{align}
&\overline {\mathtt {SLR}}_n^g = \\
&\frac{ ({\pmb \nu}^g_n)^H \mathbb E \left[ {{\mathbf h}_{n}^g} ({{\mathbf h}_{n}^g})^H \right] {\pmb \nu}^g_n } { ({\pmb \nu}^g_n)^H \mathbb E \left[ \sum\limits_{n'=1 \atop (n' \ne n)}^N    {{\mathbf h}_{n'}^g} ({{\mathbf h}_{n'}^g})^H   +  \sum\limits_{g'=1 \atop (g' \ne g)}^G \sum\limits_{n=1}^N  {{\mathbf h}_{n}^{g'}}  ({{\mathbf h}_{n}^{g'}})^H   \right]{\pmb \nu}^g_n}.
\end{align}

To gain insight into problem SL$_1$, we analyze the average channel gains on calibration errors, shown as follows
\begin{subequations}
\begin{align}
 \mathbb E \left[ {{\mathbf h}_{n}^g} ({{\mathbf h}_{n}^g})^H \right] &= \mathbb E \left[ (\mathbf{I}_{M} + \mathbf{C}^{g}_{n})\mathbf{u}^{g}_{n} ((\mathbf{I}_{M} + \mathbf{C}^{g}_{n})\mathbf{u}^{g}_{n})^H \right], 
\\
&= \mathbf{\hat U}^{g}_{n} + \mathbf{\hat U}^{g}_{n} \mathbf{\bar C}^{g}_{n},
\\
&= \mathbf{A}^{g}_{n},
\end{align}
\end{subequations}
where $\mathbf{\hat U}^{g}_{n} \in \mathbb C^{M \times M}  = {{\mathbf u}_{n}^g} ({{\mathbf u}_{n}^g})^H$, and ${\bar{\mathbf C}_{n}^g} \in \mathbb R^{M \times M} = \mathbb E \left[ \mathbf{ C}^{g}_{n} (\mathbf{ C}^{g}_{n})^H \right] = \text{diag} \left\{ \sigma^2_{cal}, \sigma^2_{cal}, ... , \sigma^2_{cal} \right\}$.

In light of the above result, the average SLR can be re-written as
\begin{align}
\overline {\mathtt {SLR}}_n^g = \frac{ ({\pmb \nu}^g_n)^H \mathbf{A}^{g}_{n} {\pmb \nu}^g_n } { ({\pmb \nu}^g_n)^H \left( \sum\limits_{n'=1 \atop (n' \ne n)}^N    \mathbf{A}^{g}_{n'}   +  \sum\limits_{g'=1 \atop (g' \ne g)}^G \sum\limits_{n=1}^N  \mathbf{A}^{g'}_{n} \right)  {\pmb \nu}^g_n}.
\end{align}

Relying on the Rayleight-Ritz quotient \cite{Parlett}, the optimal values of $\{ {\pmb \nu}^{g}_n \}$ to maximize the average SLR can be given by
\begin{align}
{\pmb \nu}^{\star g}_n = {\pmb \Phi}^g_n,
\end{align}
where ${\pmb \Phi}^g_n$ can be obtained by computing the orthonormal eigenvector corresponding to the largest eigenvalue of the matrix pair $\left(\mathbf{A}^{g}_{n},  \sum\limits_{n'=1 \atop (n' \ne n)}^N    \mathbf{A}^{g}_{n'}   +  \sum\limits_{g'=1 \atop (g' \ne g)}^G \sum\limits_{n=1}^N  \mathbf{A}^{g'}_{n}  \right)$. 

\subsection{Analysis of non-linear EH and ID performances on Non-Ideal Channel Reciprocity}
In this subsection, we present the closed-form expressions of the coverage probability of EH, and the average SINR under the calibration errors.

First, we analyze constraint \eqref{eq:MOProblemg}. In light of previous work \cite{KeXiong2017} and \eqref{eq:PCEH1}, \eqref{eq:PCEH2} and \eqref{eq:PCEH3}, the formulation regarding the non-linear EH model can be dealt with as follows
\begin{align}\label{eq:CFEH0}
&\mathcal Pr \left(\mathtt {E}_n^{g} \ge \theta_n^g \right) = \mathcal Pr \left(\sum\limits _{g'=1 }^{G} \sum\limits _{n'=1}^{N_g}\left| ({  {\mathbf h}_{n}^{g} })^T {\mathbf w}_{n'}^{g'}  \right|^2 \ge \frac{\hat \theta_n^g}{(1-\rho^g_n)}  \right), \nonumber \\
\end{align}
where
\begin{align}
\hat \theta_n^g = \mathtt{b} - \frac{1}{\mathtt{ a}}\text{ln}\left(  \frac{e^{\mathtt {ab}}({\mathtt M}^{EH} - \theta_n^g)}{e^\mathtt{ ab}\theta_n^g + \mathtt M^{EH} } \right).
\end{align}

According to \eqref{eq:theta2} and \eqref{eq:CFEH0}, constraint \eqref{eq:MOProblemg} can be represented by
\begin{align}\label{eq:CFEH1}
&{\mathcal Pr} \left( - \sum\limits _{g'=1 }^{G} \sum\limits _{n'=1}^{N_g} p_{n'}^{g'} \left| ({\pmb \nu}_{n'}^{g'})^T  {  {\mathbf h}_{n}^{g} } \right|^2 \ge \frac{\hat \theta^g_n}{{(\rho^g_n-1)}} \right)  \nonumber \\
&\le 1- \alpha^{\star g}_n.
\end{align}
It is worth reminding that $\alpha^{\star g}_n$ is the optimal value of $\alpha^{g}_n$ derived in subsection IV.A.

Indeed, the distribution of ${\mathbf h}_{n}^{g}$ under the calibration errors is $ \mathcal {CN} ({\mathbf u}_{n}^g, \sigma^2_{cal} {\hat {\mathbf U}_{n}^g}) $, where ${\hat {\mathbf U}}_{n}^g = \text{diag}\{|{\mathbf u}_{n}^g[1]|^2, ..., |{\mathbf u}_{n}^g[M]|^2 \}$. This leads to the fact that it is difficult to obtain the exact closed-form expression of \eqref{eq:CFEH1}. Additionally, the expression should be derived in a favorable formulation to the concept of the convex optimization. Therefore, this motivates us to provide an approximation of \eqref{eq:CFEH1} through {\it Lemma} 1. 
\newtheorem{lem}{Lemma}
\begin{lem}
The closed-form approximate derivation of constraint \eqref{eq:MOProblemg} can be given by
\begin{align}\label{eq:lem1}
&\pmb \varphi_n^g (\{p^g_n\}, \rho^g_n) \nonumber \\
&= {\sum\limits _{g'=1 }^{G} \sum\limits _{n'=1}^{N_g}  \sqrt{\left(  1/(1- \alpha^{\star g}_n) -1 \right){\rm var} \left[  | ({\pmb \nu}_{n'}^{g'})^T  {  {\mathbf h}_{n}^{g} } |^2 \right] } p_{n'}^{g'} } \nonumber \\
&- \sum\limits _{g'=1 }^{G} \sum\limits _{n'=1}^{N_g} \mathbb E \left[ \left| ({\pmb \nu}_{n'}^{g'})^T  {  {\mathbf h}_{n}^{g} } \right|^2 \right]p_{n'}^{g'} - \frac{\hat  \theta^g_n}{ {(\rho^g_n-1)}} \le  0.
\end{align}
\end{lem}
\begin{IEEEproof}
See Appendix A.
\end{IEEEproof}

It can be observed that the above constraint is convex over variables $\{p^g_n\}$ and $\rho^g_n$ since the term $- \frac{\hat  \theta^g_n}{ {(\rho^g_n-1)}}$ is convex (i.e. the second-order condition \cite{StephenBoyd2004}). Further, it is clear that such a closed-form derivation is a function of the expectation and variance of $| ({\pmb \nu}_{n'}^{g'})^T  {  {\mathbf h}_{n}^{g} } |^2$ for which the closed-form expressions are provided in {\it Lemma} 2.
\begin{lem}
The closed-form expressions of the expectation and variance of $\left| ({\pmb \nu}_{n'}^{g'})^T  {  {\mathbf h}_{n}^{g} } \right|^2$ can be derived respectively as follows
\begin{align}\label{eq:EXP}
\mu_{n'n}^{g'g} &= \mathbb E \left[ \left| ({\pmb \nu}_{n'}^{g'})^T  {  {\mathbf h}_{n}^{g} } \right|^2 \right] \nonumber \\
&= \text{\rm tr} \left( {\mathbf V}_{n'}^{g'} {\hat {\mathbf U}_{n}^g} \sigma^2_{cal} \right) +   ({\mathbf u}_{n}^g)^H {\mathbf V}_{n'}^{g'} {\mathbf u}_{n}^g,
\\
\upsilon_{n'n}^{g'g} &= {\rm var} \left[ \left| ({\pmb \nu}_{n'}^{g'})^T  {  {\mathbf h}_{n}^{g} } \right|^2 \right] \nonumber \\
& =  \text{\rm tr} \left( {\mathbf V}_{n'}^{g'} {\hat {\mathbf U}}_{n}^g \sigma^2_{cal} \right)^2 + 2\sigma^2_{cal} ({\mathbf u}_{n}^g)^H {\mathbf V}_{n'}^{g'}   {\hat {\mathbf U}}  {\mathbf V}_{n'}^{g'} {\mathbf u}_{n}^g.
\end{align}
\end{lem}
\begin{IEEEproof}
See Appendix B.
\end{IEEEproof}

Second, we take constraint \eqref{eq:MOProblemf} into account. On the basis of \eqref{eq:SINR}, the average SINR can be computed by {\it Lemma} 3 as follows 
\begin{lem}
According to \eqref{eq:SINR}, the closed-form derivation of the approximate average SINR can be given by
\begin{align}\label{eq:CFSINR5}
\mathbb E \left[ \mathtt {SINR}^g_n \right] 
\buildrel \Delta \over = \dfrac{ \rho^g_n \mu_{nn}^{gg} p_{n}^{g}} { \rho^g_n \left(  \bar {\mathcal I}_n^g + \sigma^2_{0} \right) + \sigma^2_{1}}.
\end{align}
where $\bar {\mathcal I}_n^g = \sum\limits _{n'=1 \atop (n' \ne n)}^{N_g} \mu_{n'n}^{gg} p_{n'}^{g} + \sum\limits _{g'=1 \atop (g' \ne g)}^{G} \sum\limits _{n'=1}^{N_g} \mu_{n'n}^{g'g} p_{n'}^{g'}$.
\end{lem}

\begin{IEEEproof}
By substituting \eqref{eq:EXP} into \eqref{eq:SINR}, we obtain \eqref{eq:CFSINR5} as in {\it Lemma} 3.
\end{IEEEproof}

\subsection{Optimal and Sub-optimal Solutions for the Problem OP$_1$ }
Since problem SL$_1$  has been relaxed, and the closed-form derivations regarding the EH and SINR metrics have been provided; problem OP$_1$ given by \eqref{eq:MOProblem} can be reformulated as
\begin{align} \label{eq:MLProblem1}
	\underset{\{p^g_n, \rho^g_n\}}\min \quad &  \sum_{g=1}^{G} \sum_{n=1}^{N_g}  p^g_n, \nonumber
\\
	 \text{s.t.:}   \quad & {\{{\mathbf w}^g_n}\} \in \mathcal F,  \quad \mathcal F \triangleq \left\{ {{\mathbf w}^g_n} |   {\mathbf w}_n^g = \sqrt{p_n^g} {\pmb \nu}_n^{\star g}, (\forall g,n)  \right\},  \nonumber \\ 
	  &{\sum\limits _{n'=1 \atop (n' \ne n)}^{N_g} \mu_{n'n}^{gg} p_{n'}^{g} + \sum\limits _{g'=1 \atop (g' \ne g)}^{G} \sum\limits _{n'=1}^{N_g} \mu_{n'n}^{g'g} p_{n'}^{g'} + \dfrac{1}{\gamma_n^g}  \mu_{nn}^{gg} p_{n}^{g}} \nonumber \\
	  &{= \sigma^2_{0}+ \dfrac{\sigma^2_{1}}{\rho^g_n}},  \quad \forall g,n,  \nonumber \\
	& \pmb \varphi_n^g (\{p^g_n\}, \rho^g_n) \le 0, \quad \forall g,n, \nonumber \\
	& {0 < \rho^g_n < 1}, \quad \forall g,n, \nonumber  \\	
	& {p^g_n \ge 0}, \quad {\forall g,n.} 
\end{align}
It can be observed that problem \eqref{eq:MLProblem1} is not convex due to the equality symbol in the second constraint. However, given the problem, the second constraint can be relaxed by replacing the equality symbol by an inequality one (i.e. $\ge$). Hence, the resulting problem becomes convex and easy to solve \cite{Gra2009}. 

In practice, since the CSI changes with time, the BS needs to update the CSI and then resolve the problem to maintain the system performance. However, this task might bring a heavily computational burden. Therefore, a sub-optimal solution that can be achieved with a much reduced complexity plays an important role. In the following, we present a method resulting in a closed-form sub-optimal solution.

First, we simplify problem \eqref{eq:MLProblem1} by setting $\rho^g_n = \rho$ $(\forall g,n)$. This implies that all users employ the same power-splitting factor. In fact, this system configuration has been adopted for multi-user SWIPT networks \cite{Ng2014}. Thus, the resulting problem of \eqref{eq:MLProblem1} can be given by
\begin{align} \label{eq:MLProblem11}
	\underset{\{p^g_n\}, \rho}\min \quad &  \sum_{g=1}^{G} \sum_{n=1}^{N_g}  p^g_n, \nonumber
\\
	 \text{s.t.:}   \quad & {\{{\mathbf w}^g_n\}} \in \mathcal F,  \quad \mathcal F \triangleq \left\{ {{\mathbf w}^g_n} |   {\mathbf w}_n^g = \sqrt{p_n^g} {\pmb \nu}_n^{\star g}, (\forall g,n)  \right \},  \nonumber \\ 
	  &{\sum\limits _{n'=1 \atop (n' \ne n)}^{N_g} \mu_{n'n}^{gg} p_{n'}^{g} + \sum\limits _{g'=1 \atop (g' \ne g)}^{G} \sum\limits _{n'=1}^{N_g} \mu_{n'n}^{g'g} p_{n'}^{g'} + \dfrac{1}{\gamma_n^g}  \mu_{nn}^{gg} p_{n}^{g}} \nonumber \\
	  &{= \sigma^2_{0}+ \dfrac{\sigma^2_{1}}{\rho}},  \quad \forall g,n,  \nonumber \\
	&\pmb \varphi_n^g (\{p^g_n\}, \rho) \le 0, \quad \forall g,n, \nonumber \\
	& {0 < \rho < 1}, \quad \forall g,n, \nonumber  \\	
	& {p^g_n \ge 0}, \quad {\forall g,n.} 
\end{align}

Second, we propose a method to facilitate the speed of solving problem \eqref{eq:MLProblem11}. Obviously, the more the variables and constraints are, the longer is the computation time required for finding solutions is. Thus, a novel transformation for problem \eqref{eq:MLProblem11} in which only one variable of $\rho$ exists is proposed.

Our idea can be explained by eliminating the variable of $\mathbf p$ $({\mathbf p} = [p_1^1 ... p_N^G]^T)$ as follows. We start with considering the second constraint of problem \eqref{eq:MLProblem11}. By representing this constraint in terms of a matrix formulation {[\cite{Ben2001}, 18.4], [\cite{Cha2009}, eq. (2)], [\cite{Ha-VuTran2017}, eq. (26)] }, the value of ${\mathbf p}$ should satisfy  
\begin{align}\label{eq:MLProblem2}
{\mathbf p} = ( {\mathbf I} - {\mathbf L} \pmb \Upsilon )^{-1} \pmb \Delta (\sigma_0^2 + \sigma_1^2\bar {\pmb \rho}),
\end{align}
where 
\begin{align}
{\mathbf L} \in \mathbb C^{NG \times NG}  = \text{diag} \{ \gamma_1^1, \ldots, \gamma_N^1, \ldots, \gamma_1^G, \ldots, \gamma_N^G \}. 
\end{align}
Moreover,
$\pmb \Upsilon \in \mathbb C^{NG \times NG}$ is a zero-diagonal matrix with 
\begin{align}
(\pmb \Upsilon)_{(k-1)G + m,(g-1)G + n} = \dfrac{\Xi^{k,g}_{m,n}}{\Xi^{k,g}_{m,m}},
\end{align}
where 
\begin{align}
\Xi^{k,g}_{m,n} = {\rm tr} \left( {\pmb \nu}_{m}^{k}({\pmb \nu}_{m}^{k})^H {\mathbf U}_{n}^g (\sigma^2_{cab} + 1) \right). 
\end{align}
In addition, $\pmb \Delta$ is a diagonal matrix, in which 
\begin{align}
{\pmb \Delta} = \text{diag} \left\{\frac{1}{\Xi^{1,1}_{1,1}}, \ldots, \frac{1}{\Xi^{g,g}_{m,m}}, \ldots, \frac{1}{\Xi^{G,G}_{M,M}} \right\} {\mathbf L}. 
\end{align}
Also, 
\begin{align}\label{eq:MLProblem2a}
\bar {\pmb \rho} = \dfrac{{\mathbf 1}_{NG}}{\rho}.
\end{align}

For convenience, we re-express \eqref{eq:MLProblem2} in terms of vector components as
\begin{align}\label{eq:MLProblem3}
p^{ g}_n = \Lambda_{g,n}({1}/{\rho}), \quad \forall g,n,
\end{align}
where function $\Lambda_{g,n}({1}/{\rho})$ can be obtained by the $((g-1)G+n)^{\text{th}}$ row of the matrix $( {\mathbf I} - {\mathbf L} \pmb \Upsilon )^{-1} \pmb \Delta (\sigma_0^2 + \sigma_1^2\bar {\pmb \rho})$.   

Moreover, since $0 < \rho < 1$, the constraint $p^{ g}_n \ge 0 \quad (\forall g,n)$ can be replaced by $( {\mathbf I} - {\mathbf L} \pmb \Upsilon )^{-1} \succeq 0$ \cite{Bambos2000}.
Now, we subsitute \eqref{eq:MLProblem3} into the problem \eqref{eq:MLProblem11} and then obtain the new formulation as
\begin{subequations}\label{eq:MLProblem4}
\begin{align}
	\underset{\rho}\min \quad &  \sum_{g=1}^{G} \sum_{n=1}^{N_g}  \Lambda_{g,n}({1}/{\rho}), 
\\
	 \text{s.t.:}\quad   
	&  \pmb \varphi_n^g (\{\Lambda_{g,n}({1}/{\rho})\}, 1/\rho) \le 0, \label{eq:MLProblem4b}\\
	& {0 < \rho < 1}, \label{eq:MLProblem4c}\\
	& {( {\mathbf I} - {\mathbf L} \pmb \Upsilon )^{-1} \succeq 0}. \label{eq:MLProblem4d}
\end{align}
\end{subequations}

Furthermore, to reduce computational burden, we aim at solving the above problem using closed-form solution through {\it Lemma 4} as follows.
\begin{lem}
Without loss of generality, constraint \eqref{eq:MLProblem4b} can be transformed into a convex form as follows
\begin{align}\label{eq:MLProblem4b2}
 (\pmb \mu_{n}^{g} - \pmb \upsilon_{n}^{g})^T{\mathbf a}\sigma_0^2 \rho^2 - \kappa_n^g \rho - (\pmb \mu_{n}^{g} - \pmb \upsilon_{n}^{g})^T{\mathbf a}\sigma_1^2 \le 0,
 \end{align}
in which,
\begin{subequations}
\begin{align}
{\mathbf a}&= ( {\mathbf I} - {\mathbf L} \pmb \Upsilon )^{-1} \pmb \Delta {\mathbf 1}_{NG}, 
\\
\pmb \mu_{n}^{g} &= [ \mu_{1n}^{1g} ... \mu_{Nn}^{Gg}]^T, 
\\
\bar {\alpha}_n^g &=  \left(  1/(1- \alpha^{\star g}_n) -1 \right), 
\\
\pmb \upsilon_{n}^{g} &= [\sqrt{\bar {\alpha}_n^g \upsilon_{1n}^{1g}} ... \sqrt{\bar {\alpha}_n^g\upsilon_{Nn}^{Gg}} ]^T, 
\\ 
\kappa_n^g &= (\pmb \mu_{n}^{g} - \pmb \upsilon_{n}^{g})^T{\mathbf a}(\sigma_0^2- \sigma_1^2) - {\hat \theta^g_n}.
\end{align}
\end{subequations}
By replacing \eqref{eq:MLProblem4b} by \eqref{eq:MLProblem4b2}, problem \eqref{eq:MLProblem4} is convex. Thus, the closed-form solution of problem \eqref{eq:MLProblem4} can be given by
\begin{align}
\rho^{\star} = \underset{{1 \le n \le N \atop 1 \le g \le G}} \min \left\{ \dfrac{ \kappa_n^g +  \sqrt{ (\kappa_n^g)^2 + 4((\pmb \mu_{n}^{g} - \pmb \upsilon_{n}^{g})^T{\mathbf a})^2\sigma_0^2\sigma_1^2  } }{2(\pmb \mu_{n}^{g} - \pmb \upsilon_{n}^{g})^T{\mathbf a}\sigma_0^2} \right\}, 
\end{align}
\end{lem}

\begin{IEEEproof}
See Appendix C.
\end{IEEEproof}

Based on the proof of {\it Lemma} 4 in Appendix C, the feasibility of problem \eqref{eq:MLProblem4} can be checked through two conditions as below. The problem is feasible if and only if
\begin{align}
( {\mathbf I} - {\mathbf L} \pmb \Upsilon )^{-1} \succeq 0,
\end{align} 
and
\begin{align}\label{eq:condi2}
\pmb \mu_{n}^{g} \ge \pmb \upsilon_{n}^{g}.
\end{align}
In practice, since the variance of calibration errors is much smaller than 1, i.e. $\sigma^2_{cal} \ll 1$, condition \eqref{eq:condi2} could be neglected for simplicity.

In summary, the principle of our method relies on exploiting the problem structure of \eqref{eq:MLProblem11} to reduce the number of variables. Specifically, the equality constraint is utilized to eliminate variable $\mathbf p$ without loss of generality. Since $\mathbf p$ can be calculated as a function of $\rho$, i.e. $\pmb \Lambda( \rho)$, the objective function and the constraints are transformed in terms of $\rho$ and $ {\pmb \Lambda}( \rho)$. In this concern, the equality constraint causing some difficulties due to its nonconvexity is also eliminated. As aforementioned, problem \eqref{eq:MLProblem4} is convex over $\rho$, thus the optimal value of $\rho$ can be conveniently obtained.

\begin{table}
\begin{centering}
\caption{Important parameters}
\par\end{centering}
\centering{}%
\begin{tabular}{|l|c|}
\hline
\textbf{Parameters} &
\textbf{System values}
\tabularnewline
\hline
Number of antennas at the BS, $M$ &
6\tabularnewline
\hline
AWGN, $\sigma^2_0$&
$10^{(-12)}$ W \tabularnewline
\hline
AWGN, $\sigma^2_1$ &
$10^{(-8)}$ W \tabularnewline
\hline
The existing plan of EH, $[q_1^1 q_2^1 q_3^1 q_1^2 q_2^2 q_1^3]$ &
[0.9 0.9 0.9 0.8 0.8 0.7]\tabularnewline
\hline
Group priorities, $[b_1^1 b_2^1 b_3^1 b_1^2 b_2^2 b_1^3]$ &
[0.1 0.1 0.1 0.2 0.2 0.3]\tabularnewline
\hline
\end{tabular}
\end{table}

\section{Numerical Results}\label{sec:NumericalResult}

In this section, the performance of the proposed resource allocation strategy is analyzed.
Without loss of generality, we assume that there exist 6 users classified into 3 groups. Specifically, group 1 includes three users (i.e. users 1, 2, and 3) located 1.5 meters far away from the BS. Group 2 consists of two users (i.e. users 4, and 5) located 3.5 meters far away from the BS. Also, group 3 contains one user whose distance to the BS is 5.5 meters (i.e. user 6). The users are randomly located in the defined ranges. Channels are assumed to have Rician distribution in which Rician factor is set to 2 dB and pathloss exponent factor is set to 2.6, i.e. in offices with soft partition \cite{Rappaport94}. Regarding the nonlinear EH model, we set $\mathtt{M}^{EH} = 24$ mW, $\mathtt{a} = 150$ and $\mathtt{b} = 0.014$ \cite{Boshkovska2017,RuihongJiang2017}. For convenience, we set the SINR thresholds as $\gamma_{n}^{g}=\gamma$, the EH thresholds as $\theta_{n}^{g}=\theta$, and the variance of calibration error as $\sigma_{cal}^2 = 0.01$. Other parameters are listed in Table I. The simulation is carried out using 1000 channel realizations. In each iteration, results are averaged on 100000 random calibration errors.

\begin{figure}[t]
\centering
\emph{\includegraphics[scale=0.5]{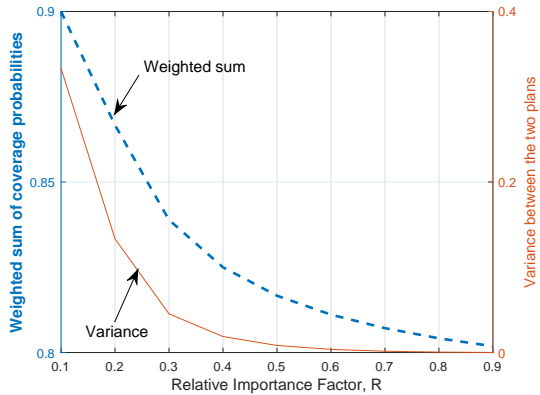}}
\caption{The trade-off between the two objectives.}
\label{fig:plan1}
\end{figure}

\begin{figure}[t]
\centering
\emph{\includegraphics[scale=0.5]{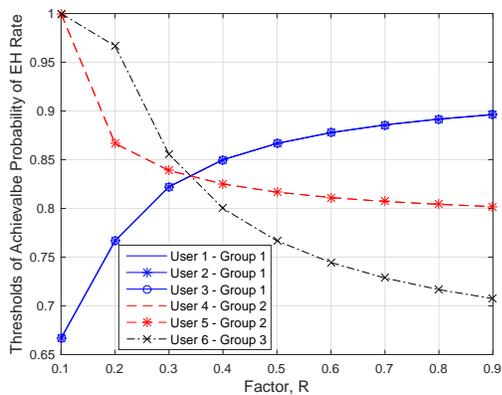}}
\caption{{Variation of the coverage probability of EH at each user}}
\label{fig:plan2}
\end{figure}

\begin{figure*}
\centering
\subfloat[EH performance of the overall system.]{\emph{\includegraphics[scale=0.5]{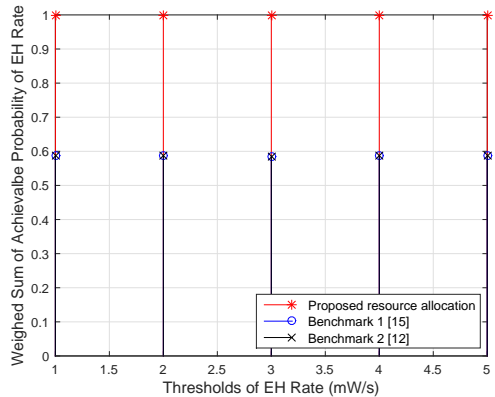}}}
\subfloat[EH performance of user 6 in group 3.]{\emph{\includegraphics[scale=0.5]{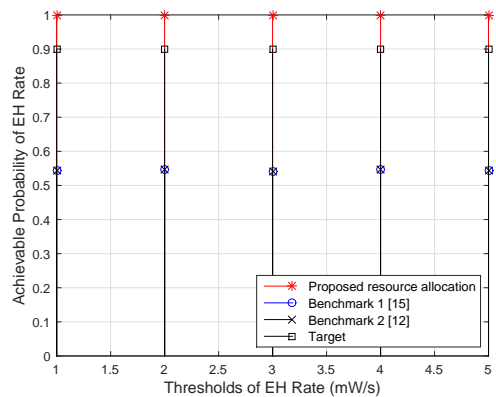}}}
\caption{EH performance enhancement ($\gamma = 2$ dB).}
\label{fig:1}
\end{figure*}

\subsection{Designing the serving plan at the upper layer}
In this work, one of the main goals is to maximize the weighted sum of the coverage probability of EH. This can be achieved by designing a new serving plan through the upper-level optimization problem OP$_2$. Thus, by using the method mentioned in subsection IV.A to solve OP$_2$, the trade-off between the two objectives is analysed and shown in Fig. \ref{fig:plan1}. As observed, increasing the factor $R$ represents that minimizing the variance between the two plans is being preferred over maximizing the weighted sum of the coverage probabilities, i.e. the higher $R$, the lower the weighted sum of the probabilities and the lower the variance. Based on this result, the network operator can select an appropriate value of $R$ for system setting according to specific network situations.

Additionally, in Fig. \ref{fig:plan2}, the varying trend for the coverage probabilities at each user in the new plan, i.e. $\{\alpha_n^g\}$, is illustrated with respect to $R$. When $R$ is close to $1$, the values of $\{\alpha_n^g\}$ are close to $\{q_n^g\}$, the original plan. When $R$ tends to $0$, this means the trend for maximizing the weighted sum of $\{\alpha_n^g\}$ becomes more important. Thus, the user with a higher priority is allocated with a higher coverage probability in the new plan. This implies that distant users may have more chances to harvest sufficient energy than nearer ones.

\subsection{Performance of the proposed resource allocation scheme}
In this section, by setting $R = 0.3$, we obtain the new plan as $[\alpha_1^{\star 1} \alpha_2^{\star 1} \alpha_3^{\star 1} \alpha_1^{\star 2} \alpha_2^{\star 2} \alpha_1^{\star 3}] = [0.8 \quad 0.8 \quad 0.8 \quad  0.85 \quad  0.85 \quad 0.9]$.

\begin{figure}[!]
\centering
\emph{\includegraphics[scale=0.5]{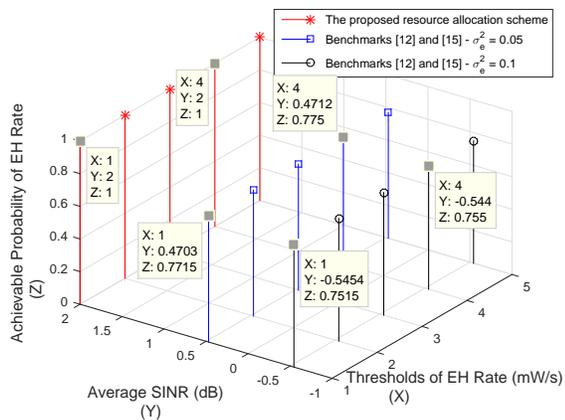}}
\caption{{The performance of EH and SINR at user 1 ($\gamma = 2$ dB).}}
\label{fig:3}
\end{figure}

Accordingly, Figs. \ref{fig:1}a-\ref{fig:1}b show the performance enhancement by a comparison between the proposed resource allocation scheme and the benchmarks \cite{QShi2014,RuihongJiang2017}. In this simulation, it is worth noting that, for a fair comparison, the scheme in benchmark \cite{QShi2014} considering a linear EH model is adopted to the nonlinear EH model. It can be seen that the two benchmarks achieve the same performance since they consider similar optimization problem and instantaneous EH and SINR constraints. As observed, the weighted sum of the probability can achieve approximately 1 and 0.6 in the cases of the proposed approach and the benchmarks, respectively. This represents a significant overall improvement of the EH performance for the considered overall system. Furthermore, in the following, we investigate the EH performance achieved at user 6, as shown in Fig. \ref{fig:1}b. Given this concern, our resource allocation scheme can guarantee the coverage probability of EH rate above the preset target. On the other hand, in the case of the benchmarks, the probability performance at user 6 is $0.54$ because the user is far from the BS. It is worth noting that the trend for the probabilities keeps constant over various EH rates. This is because the values of the probabilities mainly depend on the variance of calibration error, i.e. $\sigma_{cal}^2$, which is unchanged in this simulation. The result with different value of $\sigma_{cal}^2$ is shown in Fig. \ref{fig:3}.

\begin{figure}[!]
\centering
\emph{\includegraphics[scale=0.5]{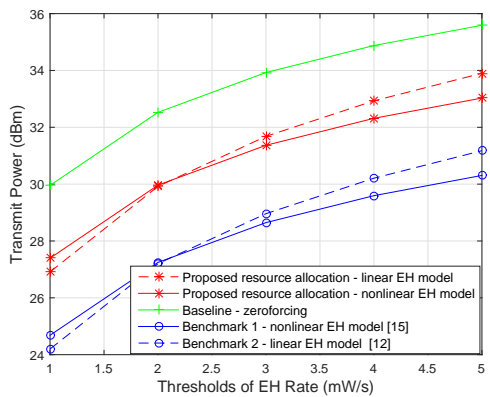}}
\caption{{A comparison between the proposed scheme and others.}}
\label{fig:comparison}
\end{figure}

In the continuity, the effectiveness of our scheme on combating the non-channel reciprocity property is shown in Fig. \ref{fig:3}. On one hand, the average SINR on the calibration error can be ensured to reach SINR threshold $\gamma$ by applying the proposed resource allocation. This can be explained by the fact that the second level problem SL$_1$ produces robust beamformers to deal with the non-channel reciprocity property efficiently. On the other hand, the SINR performance does not meet the requirement by employing the benchmarks. The more the SINR is reduced when the more the calibration error is considered. Based on Figs. \ref{fig:1}a, \ref{fig:1}b and \ref{fig:3}, it can be concluded that not only the user fairness but the effect of the non-ideal channel reciprocity can also be properly managed.

Fig. \ref{fig:comparison} provides a comparison in terms of the transmit power between the proposed resource allocation scheme, the existing benchmarks and a baseline. To highlight the effect of nonlinear and linear EH models, we show the transmit power for the proposed scheme and the benchmarks \cite{QShi2014,RuihongJiang2017}.  Note that the energy conversion effciency is set to $0.5$ when considering the linear EH model. Here, we can evaluate the mismatch between the linear EH model and the practical nonlinear EH model. In particular, it is observed that the proposed scheme sacrifices {$2.7$} dB more than the benchmarks in both of the models to maintain the required ID and EH performances. Further, a comparison with a baseline that consists of the well-known zero-forcing beamformer is shown. A gap of approximately {$2.9$} dB between the baseline and the proposed resource allocation scheme exists. This is because the interference caused by the SLR beamformer can be useful for EH performance. So, the proposed scheme requires less transmit power to achieve the same targeted
system performance than the baseline does.

\begin{figure}
\centering
\emph{\includegraphics[scale=0.5]{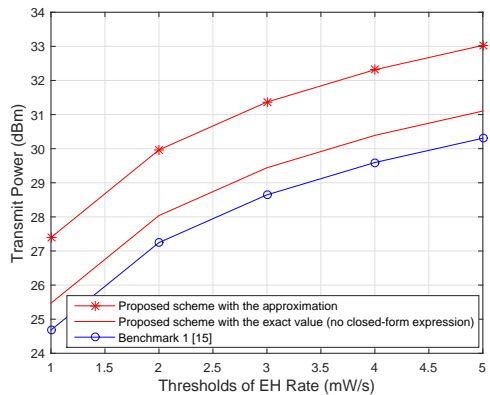}}
\caption{{Performance with the approximations.}}
\label{fig:approximation}
\end{figure}

Additionally, the closed-form approximate expressions of the average SINR and the coverage probability of EH, shown in {\it Lemmas} $1$ and $3$, play an important role in the performance of the proposed scheme. In Fig. \ref{fig:approximation}, the tightness of the approximations is verified. It is evaluated that there exists a gap of approximately {$1.9$} dB regarding the transmit power. Since the exact closed-form expressions are not available, using the derived approximation can be an efficient choice. 

\begin{figure}[t]
\centering
\emph{\includegraphics[scale=0.5]{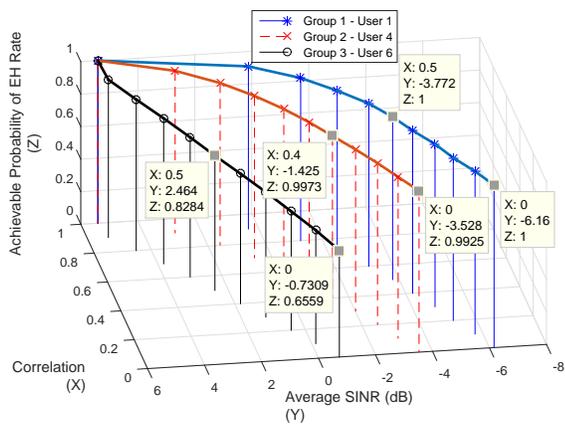}}
\caption{{Performance loss of the average SINR and the EH.}}
\label{fig:4}
\end{figure}

\begin{figure}[!]
\centering
\emph{\includegraphics[scale=0.5]{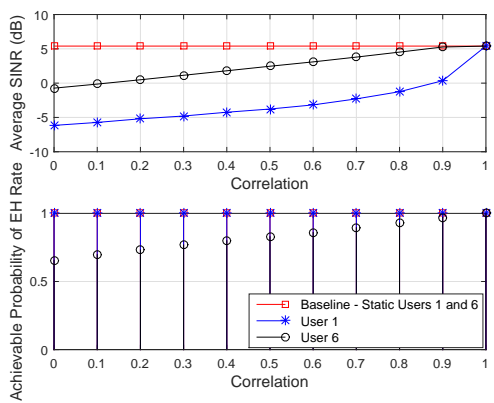}}
\caption{{Performance loss at near and far users.}}
\label{fig:42}
\end{figure}

\subsection{Performance loss due to user mobility, and the suboptimal solution}

This part of the simulation is to investigate performance loss due to the user mobility, and then highlight the role of the suboptimal solution.
Assuming that users 1, 4, and 6 are moving, Fig. \ref{fig:4} depicts the system performance loss due to the user mobility in each group. In the simulation and for the sake of convenience, we consider a mobility scenario in which each user moves to a new random location keeping its distance to the BS unchanged. In other words, the corresponding propagation loss and the assigned user group are unaffected. Particularly, since the actual CSI is unknown at the BS, this inspires us to simulate user mobility using a correlation model. In this concern, the difference between the actual CSI and the estimated one at the BS can be measured by correlation coefficients.  Thus, the simulation can be carried out by generating CSI correlated with the estimated CSI. In fact, the motion of the users can lead the estimated CSI at the BS to be outdated promptly. The latter results in a non-trivial reduction of the SINR and EH performances. On this basis, Figs. \ref{fig:4} and \ref{fig:42} indicate the average SINR degrade with respect to the correlation where the smaller correlation implies the higher user mobility and yields the higher average SINR loss. Further, Figs. \ref{fig:4} and \ref{fig:42} illustrate a reduction of the achievable coverage probability of EH rate at each user. Especially, it seems that the EH performance at user 1 unchanged even when there is no correlation between the actual and estimated CSI. Besides, it can be observed that the shorter the distance between the users and the BS is, the worse the SINR performance is at each moving user. This is because nearer users are exposed by denser interferences compared to farther users. 

\begin{figure}
\centering
{\emph{\includegraphics[scale=0.5]{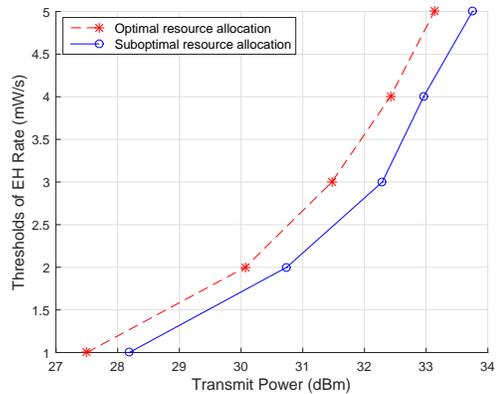}}}
\caption{{A comparison between the optimal and suboptimal solutions ($\gamma$ = 5).}}
\label{fig:5}
\end{figure}

As shown in Fig. \ref{fig:4}, the system performance decreases significantly due to the user mobility. When the users move, the BS should update the CSI and then re-compute the power vectors, the beamformers and the power-splitting factors. This task might bring a heavily computational burden to the BS, especially in the case where the user mobility is high. Thus, the proposed suboptimal solution for which its closed-form derivation provided in this work can be very well considered as an efficient alternative approach. In this concern, Fig. \ref{fig:5} shows a performance comparison between the optimal and suboptimal solutions. On this basis, it can be evaluated that there exists a power gap of approximate {$0.75$} dBm between two approaches. Indeed, using the suboptimal approach implies sacrificing an additional amount of transmit power, however, this also significantly reduces computational burden.

\section{Conclusion}\label{sec:Conclussion}
In this paper, we propose a novel approach to allocate resources for the SWIPT system under a nonlinear EH model, in which multiple critical requirements, such as minimizing transmit power, maximizing the weighted sum of coverage probability of EH and minimizing the effect of calibration error on the ID performance, need to be concurrently managed. Such requirements are mathematically structured on the cross-layer multi-level formulation where the upper-layer and physical layer problems are devised. On this basis, we provide the algorithms and the closed-form solutions to tackle these problems. The obtained numerical results show that not only the user fairness but the effect of the non-ideal channel reciprocity can also be properly managed using the proposed resource allocation scheme. A comparison between the proposed scheme, the existing benchmarks, and the baseline has been provided to highlight the effectiveness of our work. Particularly, the results imply that when users are moving, farther users can obtain a better SINR performance than nearer users. This is because the nearer users are more susceptible to interferences. To this end, the suboptimal solution can be considered as an efficient alternative to maintain the system performance in the case of fast time-varying channels.

\appendices
\section{Proof of Lemma 1}
To compute an approximation for \eqref{eq:CFEH1}, we begin with considering the simplified form (setting $G=N_1=1$) as follows
\begin{align}\label{eq:CFEH2} 
{\mathcal Pr} \left( - p_{1}^{1} \left| ({\pmb \nu}_{1}^{1})^T  {  {\mathbf h}_{1}^{1} } \right|^2 \ge \frac{\hat \theta^1_1}{ {(\rho^1_1-1)}} \right)  \le 1- \alpha^{\star 1}_1.
\end{align}

Following the one-sided Chebyshev-inequality, we have
\begin{align}\label{eq:CFEH3}
&{\mathcal Pr} \left( - p_{1}^{1} \left| ({\pmb \nu}_{1}^{1})^T  {  {\mathbf h}_{1}^{1} } \right|^2 + \mathbb E \left[ p_{1}^{1} \left| ({\pmb \nu}_{1}^{1})^T  {  {\mathbf h}_{1}^{1} } \right|^2 \right]  \ge t \right) \nonumber \\ 
&\le \frac{1}{1 + t^2/\text{var} \left[ p_{1}^{1}\left| ({\pmb \nu}_{1}^{1})^T  {  {\mathbf h}_{1}^{1} } \right|^2 \right]},
\end{align}
where $t \ge 0$. By taking
\begin{align}\label{eq:CFEH3-1} 
t = \dfrac{\hat \theta^1_1}{{(\rho^1_1-1)}} + \mathbb E \left[ p_{1}^{1} \left| ({\pmb \nu}_{1}^{1})^T  {  {\mathbf h}_{1}^{1} } \right|^2 \right], 
\end{align}
 one can see that the expression \eqref{eq:CFEH2} is equivalent to the one \eqref{eq:CFEH3}. On this basis, we obtain
\begin{align}\label{eq:CFEH4}
&{\mathcal Pr} \left( - p_{1}^{1} \left| ({\pmb \nu}_{1}^{1})^T  {  {\mathbf h}_{1}^{1} } \right|^2 \ge \frac{\hat \theta^1_1}{ {(\rho^1_1-1)}} \right) \nonumber \\
&\le \frac{\text{var} \left[ p_{1}^{1} \left| ({\pmb \nu}_{1}^{1})^T  {  {\mathbf h}_{1}^{1} } \right|^2 \right]}{\text{var} \left[ p_{1}^{1} \left| ({\pmb \nu}_{1}^{1})^T  {  {\mathbf h}_{1}^{1} } \right|^2 \right] + \left( \dfrac{\theta^1_1}{\xi^1_1 {(\rho^1_1-1)}} + \mathbb E \left[ p_{1}^{1} \left| ({\pmb \nu}_{1}^{1})^T  {  {\mathbf h}_{1}^{1} } \right|^2 \right] \right)^2}.
\end{align}

As a result, the inequality
\begin{align}\label{eq:CFEH5}
&\frac{\text{var} \left[ p_{1}^{1} \left| ({\pmb \nu}_{1}^{1})^T  {  {\mathbf h}_{1}^{1} } \right|^2 \right]}{\text{var} \left[ p_{1}^{1} \left| ({\pmb \nu}_{1}^{1})^T  {  {\mathbf h}_{1}^{1} } \right|^2 \right] + \left( \dfrac{\hat \theta^1_1}{ {(\rho^1_1-1)}} + \mathbb E \left[ p_{1}^{1} \left| ({\pmb \nu}_{1}^{1})^T  {  {\mathbf h}_{1}^{1} } \right|^2 \right] \right)^2} \nonumber
\\
&\le 1- {\alpha^{\star 1}_1}.
\end{align}
implies the probability inequality given in \eqref{eq:CFEH2}.

Additionally, since $t \ge 0$, the expression in \eqref{eq:CFEH5} can be re-written by
\begin{align}\label{eq:CFEH6}
&\sqrt{\text{var} \left[ p_{1}^{1} \left| ({\pmb \nu}_{1}^{1})^T  {  {\mathbf h}_{1}^{1} } \right|^2 \right] \left(  1/(1- \alpha^{\star 1}_1) -1 \right)} \nonumber \\
&\le \mathbb E \left[ p_{1}^{1} \left| ({\pmb \nu}_{1}^{1})^T  {  {\mathbf h}_{1}^{1} } \right|^2 \right]  + \dfrac{\hat \theta^1_1}{ {(\rho^1_1-1)}}.
\end{align}

By generalizing \eqref{eq:CFEH2}-\eqref{eq:CFEH6}, 
\begin{align}\label{eq:CFEH7}
&  \sqrt{\left(  1/(1- \alpha^{\star g}_n) -1 \right) {\sum\limits _{g'=1 }^{G} \sum\limits _{n'=1}^{N_g} {\rm var} \left[  | ({\pmb \nu}_{n'}^{g'})^T  {  {\mathbf h}_{n}^{g} } |^2 \right] } (p_{n'}^{g'})^2 } \nonumber \\
& \le \sum\limits _{g'=1 }^{G} \sum\limits _{n'=1}^{N_g} \mathbb E \left[ \left| ({\pmb \nu}_{n'}^{g'})^T  {  {\mathbf h}_{n}^{g} } \right|^2 \right]p_{n'}^{g'} + \frac{\hat \theta^g_n}{ {(\rho^g_n-1)}}.
\end{align}
Also, it is known that
\begin{align}
 &\sqrt{ {\sum\limits _{g'=1 }^{G} \sum\limits _{n'=1}^{N_g} {\rm var} \left[  | ({\pmb \nu}_{n'}^{g'})^T  {  {\mathbf h}_{n}^{g} } |^2 \right] } (p_{n'}^{g'})^2 } \nonumber \\
&\le {\sum\limits _{g'=1 }^{G} \sum\limits _{n'=1}^{N_g}  \sqrt{{\rm var} \left[  | ({\pmb \nu}_{n'}^{g'})^T  {  {\mathbf h}_{n}^{g} } |^2 \right] } p_{n'}^{g'} }.
\end{align}
Thus, \eqref{eq:CFEH7} can be approximated as follows
\begin{align}
&{\sum\limits _{g'=1 }^{G} \sum\limits _{n'=1}^{N_g}  \sqrt{\left(  1/(1- \alpha^{\star g}_n) -1 \right){\rm var} \left[  | ({\pmb \nu}_{n'}^{g'})^T  {  {\mathbf h}_{n}^{g} } |^2 \right] } p_{n'}^{g'} } \nonumber \\
& \le \sum\limits _{g'=1 }^{G} \sum\limits _{n'=1}^{N_g} \mathbb E \left[ \left| ({\pmb \nu}_{n'}^{g'})^T  {  {\mathbf h}_{n}^{g} } \right|^2 \right]p_{n'}^{g'} + \frac{\hat \theta^g_n}{ {(\rho^g_n-1)}}.
\end{align}
This completes our proof.

\section{Proof of Lemma 2}
To calculate $\mathbb E \left[ \left| ({\pmb \nu}_{n'}^{g'})^T  {  {\mathbf h}_{n}^{g} } \right|^2 \right]$, it is useful to represent it into a formulation regarding the trace operator as follows
\begin{align}
\mathbb E \left[ \left| ({\pmb \nu}_{n'}^{g'})^T  {  {\mathbf h}_{n}^{g} } \right|^2 \right] = \mathbb E \left[ \text{tr} \left(    {\mathbf V}_{n'}^{g'}   {  {\mathbf h}_{n}^{g} } ({\mathbf h}_{n}^{g})^H  \right) \right].
\end{align}

Due to the facts that (i) both $\mathbb E \left[ . \right]$ and tr are linear operators and (ii) ${\mathbf h}_{n}^g \sim \mathcal {CN} ({\mathbf u}_{n}^g, \sigma^2_{cal} {{\mathbf U}_{n}^g}) $, where ${{{\mathbf U}}}_{n}^g = \text{diag}\{|{\mathbf u}_{n}^g[1]|^2, ..., |{\mathbf u}_{n}^g[M]|^2 \}$, we can derive
\begin{align}
&\mathbb E \left[ \text{tr} \left(   {\mathbf V}_{n'}^{g'}   {  {\mathbf h}_{n}^{g} } ({\mathbf h}_{n}^{g})^H  \right) \right] =  \text{tr} \left(    {\mathbf V}_{n'}^{g'}   \mathbb E \left[{  {\mathbf h}_{n}^{g} } ({\mathbf h}_{n}^{g})^H  \right] \right) \nonumber \\
&= \text{tr} \left( {\mathbf V}_{n'}^{g'} {\hat {\mathbf U}_{n}^g} \sigma^2_{cal} \right) +   ({\mathbf u}_{n}^g)^H {\mathbf V}_{n'}^{g'} {\mathbf u}_{n}^g.
\end{align}

On the other hand, one can observe that computing the variance of $\left| ({\pmb \nu}_{n'}^{g'})^T  {  {\mathbf h}_{n}^{g} } \right|^2$ is quite complicated. Fortunately, based on the work in [Chapter 2, \cite{Searle}], after some manipulations, the cumulant generating function (CGF) of $\left| ({\pmb \nu}_{n'}^{g'})^T  {  {\mathbf h}_{n}^{g} } \right|^2$ can be given by
\begin{align}
&\mathtt{CGF}_k \left( | ({\pmb \nu}_{n'}^{g'})^T  {  {\mathbf h}_{n}^{g} } |^2 \right) =  (k-1)! \left[ \text{tr} \left( {\mathbf V}_{n'}^{g'} {\hat {\mathbf U}}_{n}^g \sigma^2_{cal} \right)^k \right. \nonumber \\
&\left. + k ({\mathbf u}_{n}^g)^H {\mathbf V}_{n'}^{g'} \left( \sigma^2_{cal} {\hat {\mathbf U}}_{n}^g   {\mathbf V}_{n'}^{g'} \right)^{k-1} {\mathbf u}_{n}^g  \right].
\end{align}

Particularly, by setting $k = 2$, the variance of $\left| ({\pmb \nu}_{n'}^{g'})^T  {  {\mathbf h}_{n}^{g} } \right|^2$ can be shown as
\begin{align}
\text{var} \left[ \left| ({\pmb \nu}_{n'}^{g'})^T  {  {\mathbf h}_{n}^{g} } \right|^2 \right] &= \text{tr} \left( {\mathbf V}_{n'}^{g'} {\hat {\mathbf U}}_{n}^g \sigma^2_{cal} \right)^2 \nonumber \\
&+ 2\sigma^2_{cal} ({\mathbf u}_{n}^g)^H {\mathbf V}_{n'}^{g'}   {\hat {\mathbf U}}  {\mathbf V}_{n'}^{g'} {\mathbf u}_{n}^g.
\end{align}
This completes our proof.

\section{Proof of Lemma 4}
First, we consider constraint \eqref{eq:MLProblem4b}.
According to \eqref{eq:lem1}, the constraint can be re-expressed as
\begin{align}\label{eq:x}
(\pmb \mu_{n}^{g} - \pmb \upsilon_{n}^{g})^T{\mathbf a}(\sigma_0^2 + \frac{\sigma_1^2}{\rho})  \ge \frac{\hat \theta^g_n}{ {(1-\rho)}},
\end{align}
where 
\begin{subequations}
\begin{align}
{\mathbf a} &= ( {\mathbf I} - {\mathbf L} \pmb \Upsilon )^{-1} \pmb \Delta {\mathbf 1}_{NG},
\\
\pmb \mu_{n}^{g} &= [ \mu_{1n}^{1g} ... \mu_{Nn}^{Gg}]^T, 
\\
\bar {\alpha}_n^g &=  \left(  1/(1- \alpha^{\star g}_n) -1 \right),
\\
\pmb \upsilon_{n}^{g} &= [\sqrt{\bar {\alpha}_n^g \upsilon_{1n}^{1g}} ... \sqrt{\bar {\alpha}_n^g\upsilon_{Nn}^{Gg}} ]^T.
\end{align}
\end{subequations}
By inverting both sides, we can obtain the derivation as below
\begin{align}
 \frac{\rho }{\rho(\pmb \mu_{n}^{g} - \pmb \upsilon_{n}^{g})^T{\mathbf a}\sigma_0^2 + (\pmb \mu_{n}^{g} - \pmb \upsilon_{n}^{g})^T{\mathbf a}\sigma_1^2}  \le \frac{ {(1-\rho)}}{\hat \theta^g_n}.
\end{align}
After some manipulations, the constraint can be further transformed into a second-order inequation as
\begin{align}\label{eq:y}
(\pmb \mu_{n}^{g} - \pmb \upsilon_{n}^{g})^T{\mathbf a}\sigma_0^2 \rho^2 - \kappa_n^g \rho - (\pmb \mu_{n}^{g} - \pmb \upsilon_{n}^{g})^T{\mathbf a}\sigma_1^2 \le 0,
\end{align}
in which, 
\begin{align}
\kappa_n^g = (\pmb \mu_{n}^{g} - \pmb \upsilon_{n}^{g})^T{\mathbf a}(\sigma_0^2- \sigma_1^2) - {\hat \theta^g_n}.
\end{align}
{In particular, according to \eqref{eq:x}, it can be induced that $(\pmb \mu_{n}^{g} - \pmb \upsilon_{n}^{g})^T{\mathbf a} \ge 0$ to make the problem feasible. This yields that \eqref{eq:y} is a convex quadratic constraint \cite{StephenBoyd2004}. In other words, constraint \eqref{eq:MLProblem4b} has been transformed into a convex formulation.} 

Second, we analyze the convexity of the objective function of problem \eqref{eq:MLProblem4}. 
Taking \eqref{eq:MLProblem2}, \eqref{eq:MLProblem2a}, and \eqref{eq:MLProblem3} into account, it can be evaluated that since matrix $( {\mathbf I} - {\mathbf L} \pmb \Upsilon )^{-1}\pmb \Delta$ is positive element-wise \cite{Bambos2000}, $\Lambda_{g,n}({1}/{\rho})$ is convex. Thus, the objective function of problem \eqref{eq:MLProblem4} can be seen as a summation of convex functions. Therefore, the objective function is convex.

Based on the above discussion, it can be concluded that problem \eqref{eq:MLProblem4} is convex.

Next, since the objective function of problem \eqref{eq:MLProblem4} is non-increasing, the closed-form solution can be given by
\begin{align}
\rho^{\star} =  \underset{{1 \le n \le N \atop 1 \le g \le G}}   \min \left\{ \dfrac{ \kappa_n^g +  \sqrt{ (\kappa_n^g)^2 + 4((\pmb \mu_{n}^{g} - \pmb \upsilon_{n}^{g})^T{\mathbf a})^2\sigma_0^2\sigma_1^2  } }{2(\pmb \mu_{n}^{g} - \pmb \upsilon_{n}^{g})^T{\mathbf a}\sigma_0^2} \right\}.
\end{align}
This completes our proof.

\bibliographystyle{IEEEtran}
\bibliography{IEEEabrv,REF}

\begin{thebibliography}{10}
\providecommand{\url}[1]{#1}
\csname url@samestyle\endcsname
\providecommand{\newblock}{\relax}
\providecommand{\bibinfo}[2]{#2}
\providecommand{\BIBentrySTDinterwordspacing}{\spaceskip=0pt\relax}
\providecommand{\BIBentryALTinterwordstretchfactor}{4}
\providecommand{\BIBentryALTinterwordspacing}{\spaceskip=\fontdimen2\font plus
\BIBentryALTinterwordstretchfactor\fontdimen3\font minus
  \fontdimen4\font\relax}
\providecommand{\BIBforeignlanguage}[2]{{%
\expandafter\ifx\csname l@#1\endcsname\relax
\typeout{** WARNING: IEEEtran.bst: No hyphenation pattern has been}%
\typeout{** loaded for the language `#1'. Using the pattern for}%
\typeout{** the default language instead.}%
\else
\language=\csname l@#1\endcsname
\fi
#2}}
\providecommand{\BIBdecl}{\relax}
\BIBdecl

\bibitem{Gupta2015}
A.~Gupta and R.~K. Jha, ``A survey of {5G} network: Architecture and emerging
  technologies,'' \emph{IEEE Access}, vol.~3, pp. 1206 -- 1232, July 2015.

\bibitem{Ekram2015}
E.~Hossain and M.~Hasan, ``{5G} cellular: Key enabling technologies and
  research challenges,'' \emph{{IEEE} Instrum. Meas. Mag.}, vol.~18, no.~3, pp.
  11--21, June 2015.

\bibitem{StefanoBuzzi2016}
S.~Buzzi, C.-L. I, T.~E. Klein, H.~V. Poor, C.~Yang, and A.~Zappone, ``A survey
  of energy-efficient techniques for {5G} networks and challenges ahead,''
  \emph{{IEEE} J. Sel. Areas Commun.}, vol.~34, no.~4, pp. 697--709, April
  2016.

\bibitem{Grover2010}
P.~Grover and A.~Sahai, ``{Shannon} meets tesla: Wireless information and power
  transfer,'' in \emph{IEEE International Symposium on Information Theory
  Proceedings (ISIT)}, Austin, TX, USA, June 2010, pp. 2363--2367.

\bibitem{Lu2015}
X.~Lu, P.~Wang, D.~Niyato, D.~I. Kim, and Z.~Han, ``Wireless networks with {RF}
  energy harvesting: A contemporary survey,'' \emph{IEEE Comm. Surveys \&
  Tutorials}, vol.~17, no.~2, pp. 757--789, 2015.

\bibitem{DusitNiyato}
D.~Niyato, D.~I. Kim, M.~Maso, and Z.~Han, ``Wireless powered communication
  networks: Research directions and technological approaches,'' \emph{{IEEE}
  Wireless Commun. Mag.}, vol.~24, no.~6, pp. 88 -- 97, December 2017.

\bibitem{HaVuTranPotentials}
H.-V. Tran and G.~Kaddoum, ``{RF} wireless power transfer: Regreening the
  future networks,'' \emph{IEEE Potentials}, vol.~37, no.~2, pp. 35 -- 41,
  March-April 2018.

\bibitem{SWIPTKaddoum}
G.~Kaddoum, H.-V. Tran, L.~Kong, and M.~Atalla, ``Design of simultaneous
  wireless information and power transfer scheme for short reference {DCSK}
  communication systems,'' \emph{{IEEE} Trans. Commun.}, vol.~65, no.~1, pp.
  431 -- 443, Jan. 2017.

\bibitem{TuanAnhLe}
T.~A. Le, Q.-T. Vien, H.~X. Nguyen, D.~W.~K. Ng, and R.~Schober, ``Robust
  chance-constrained optimization for power-efficient and secure {SWIPT}
  systems,'' \emph{IEEE Transactions on Green Communications and Networking},
  vol.~1, no.~3, pp. 333 -- 346, May 2017.

\bibitem{PhamVietTuan}
P.~V. Tuan and I.~Koo, ``Optimal multiuser {MISO} beamforming for
  power-splitting {SWIPT} cognitive radio networks,'' \emph{IEEE Access},
  vol.~5, pp. 14\,141 -- 14\,153, July 2017.

\bibitem{Zhang2013}
R.~Zhang and C.~K. Ho, ``{MIMO} broadcasting for simultaneous wireless
  information and power transfer,'' \emph{{IEEE} Trans. Wireless Commun.},
  vol.~12, no.~5, pp. 1989--2001, 2013.

\bibitem{QShi2014}
Q.~Shi, L.~Liu, W.~Xu, and R.~Zhang, ``Joint transmit beamforming and receive
  power splitting for {MISO} {SWIPT} systems,'' \emph{{IEEE} Trans. Wireless
  Commun.}, vol.~13, no.~6, pp. 3269--3280, June 2014.

\bibitem{Boshkovska2015a}
E.~Boshkovska, D.~W.~K. Ng, N.~Zlatanov, and R.~Schober, ``Practical non-linear
  energy harvesting model and resource allocation for {SWIPT} systems,''
  \emph{{IEEE} Commun. Lett.}, vol.~19, no.~12, pp. 2082 -- 2085, Dec. 2015.

\bibitem{Boshkovska2017}
E.~Boshkovska, D.~W.~K. Ng, N.~Zlatanov, A.~Koelpin, and R.~Schober, ``Robust
  resource allocation for {MIMO} wireless powered communication networks based
  on a non-linear {EH} model,'' \emph{{IEEE} Trans. Commun.}, vol.~65, no.~6,
  pp. 1984 -- 1999, February 2017.

\bibitem{RuihongJiang2017}
R.~Jiang, K.~Xiong, P.~Fan, Y.~Zhang, and Z.~Zhong, ``Optimal design of {SWIPT}
  systems with multiple heterogeneous users under non-linear energy harvesting
  model,'' \emph{IEEE Access}, vol.~5, pp. 11\,479 -- 11\,489, June 2017.

\bibitem{KeXiong2017}
K.~Xiong, B.~Wang, and K.~J.~R. Liu, ``Rate-energy region of {SWIPT} for {MIMO}
  broadcasting under nonlinear energy harvesting model,'' \emph{{IEEE} Trans.
  Wireless Commun.}, vol.~16, no.~8, pp. 5147 -- 5161, May 2017.

\bibitem{HJu2014}
H.~Ju and R.~Zhang, ``Throughput maximization for wireless powered
  communication networks,'' \emph{IEEE Trans. Wireless Commun.}, vol.~13,
  no.~1, pp. 418--428, Jan 2014.

\bibitem{Chingoska2016}
H.~Chingoska, Z.~Hadzi-Velkov, I.~Nikoloska, and N.~Zlatanov, ``Resource
  allocation in wireless powered communication networks with non-orthogonal
  multiple access,'' \emph{IEEE Wireless Comm. Lett.}, vol.~5, no.~6, pp. 684
  -- 687, December 2016.

\bibitem{PDD17}
P.~D. Diamantoulakis and G.~K. Karagiannidis, ``Maximizing proportional
  fairness in wireless powered communications,'' \emph{IEEE Wireless Comm.
  Lett.}, vol.~6, no.~2, pp. 202 -- 205, April 2017.

\bibitem{PDD171}
P.~D. Diamantoulakis, K.~N. Pappi, G.~K. Karagiannidis, H.~Xing, and
  A.~Nallanathan, ``Joint downlink/uplink design for wireless powered networks
  with interference,'' \emph{IEEE Access}, vol.~5, pp. 1534 -- 1547, January
  2017.

\bibitem{Liu2014}
L.~Liu, R.~Zhang, and K.-C. Chua, ``Multi-antenna wireless powered
  communication with energy beamforming,'' \emph{{IEEE} Trans. Commun.},
  vol.~62, no.~12, pp. 4349--4361, 2014.

\bibitem{Tabassum2015}
H.~Tabassum, E.~Hossain, A.~Ogundipe, and D.~I. Kim, ``Wireless-powered
  cellular networks: Key challenges and solution techniques,'' \emph{{IEEE}
  Commun. Mag.}, vol.~53, no.~6, pp. 63 -- 71, June 2015.

\bibitem{Mishra2017}
D.~Mishra, S.~De, and D.~Krishnaswamy, ``Dilemma at {RF} energy harvesting
  relay: Downlink energy relaying or uplink information transfer?''
  \emph{{IEEE} Trans. Wireless Commun.}, vol.~16, no.~8, pp. 4939 -- 4955, May
  2017.

\bibitem{EmilBjornson2014}
E.~Bjornson, E.~A. Jorswieck, M.~Debbah, and B.~Ottersten, ``Multiobjective
  signal processing optimization: The way to balance conflicting metrics in
  {5G} systems,'' \emph{{IEEE} Signal Process. Mag.}, vol.~31, no.~6, pp. 14 --
  23, November 2014.

\bibitem{Marler2004}
R.~Marler and J.~Arora, ``Survey of multi-objective optimization methods for
  engineering,'' \emph{Structural and Multidisciplinary Optimization}, vol.~26,
  pp. 369--395, 2004.

\bibitem{DerrickNg2016}
D.~W.~K. Ng, E.~S. Lo, and R.~Schober, ``Multiobjective resource allocation for
  secure communication in cognitive radio networks with wireless information
  and power transfer,'' \emph{{IEEE} Trans. Veh. Commun.}, vol.~65, no.~5, pp.
  3166 -- 3184, May 2016.

\bibitem{YanSun2016}
Y.~Sun, D.~W.~K. Ng, J.~Zhu, and R.~Schober, ``Multi-objective optimization for
  robust power efficient and secure full-duplex wireless communication
  systems,'' \emph{{IEEE} Trans. Wireless Commun.}, vol.~15, no.~8, pp. 5511 --
  5526, August 2016.

\bibitem{Gra2009}
M.~Grant and S.~Boyd, \emph{CVX: Matlab software for disciplined convex
  programming}, June 2009.

\bibitem{3GPP091794}
{3GPP}, ``Hardware calibration requirement for dual layer beamforming,''
  Huawei, Tech. Rep. R1-091794, 2009.

\bibitem{Vieira14}
J.~Vieira, F.~Rusek, and F.~Tufvesson, ``Reciprocity calibration methods for
  massive {MIMO} based on antenna coupling,'' in \emph{IEEE Global
  Communications Conference}, Austin, TX, USA, December 2014.

\bibitem{Ng2014}
D.~Ng, E.~Lo, and R.~Schober, ``Robust beamforming for secure communication in
  systems with wireless information and power transfer,'' \emph{{IEEE} Trans.
  Wireless Commun.}, vol.~13, no.~8, pp. 4599--4615, 2014.

\bibitem{GrahamUpton}
G.~Upton and I.~Cook, \emph{A Dictionary of Statistics}, 3rd~ed.\hskip 1em plus
  0.5em minus 0.4em\relax Oxford University Press, 2014.

\bibitem{EBjornson2013}
E.~Bjornson and E.~Jorswieck, ``Optimal resource allocation in coordinated
  multi-cell systems,'' \emph{Foundations and Trends in Communications and
  Information Theory}, vol.~9, no. 2-3, pp. 113--381, 2013.

\bibitem{Caramia2008}
M.~Caramia and P.~Dell'Olmo, \emph{Multi-objective management in freight
  logistics: Increasing capacity, service level and safety with optimization
  algorithms}.\hskip 1em plus 0.5em minus 0.4em\relax Springer Science \&
  Business Media, 2008.

\bibitem{Ge2011}
D.~Ge, X.~Jiang, and Y.~Ye, ``A note on the complexity of lp minimization,''
  \emph{Mathematical Programming}, vol. 129, no.~2, pp. 285--299, 2011.

\bibitem{Baraniuk2007}
R.~Baraniuk, ``Compressive sensing,'' \emph{{IEEE} Signal Process. Mag.},
  vol.~24, no.~4, pp. 118 -- 121, August 2007.

\bibitem{FenChen2013}
F.~Chen and Y.~Zhang, ``Sparse hyperspectral unmixing based on constrained lp -
  l2 optimization,'' \emph{{IEEE} Geosci. Remote Sens. Lett.}, vol.~10, no.~6,
  pp. 1142 -- 1146, September 2013.

\bibitem{Iordache2014}
M.-D. Iordache, J.~M. Bioucas-Dias, and A.~Plaza, ``Collaborative sparse
  regression for hyperspectral unmixing,'' \emph{{IEEE} Trans. Geosci. Remote
  Sens.}, vol.~12, no.~1, pp. 341 -- 354, January 2014.

\bibitem{DAgarwal}
D.~Agarwal, B.-C. Chen, P.~Elango, and X.~Wang, ``Personalized click shaping
  through lagrangian duality for online recommendation,'' in \emph{the 35th
  International {ACM} {SIGIR} Conference on Research and Development in
  Information Retrieval}, Portland, Oregon, USA, August 2012.

\bibitem{StephenBoyd2004}
S.~Boyd and L.~Vandenberghe, \emph{Convex Optimization}.\hskip 1em plus 0.5em
  minus 0.4em\relax Cambridge University Press, 2004.

\bibitem{Sha2007}
F.~Sha, Y.~Lin, L.~K. Saul, and D.~D. Lee, ``Multiplicative updates for
  nonnegative quadratic programming,'' \emph{Journal Neural Computation},
  vol.~19, no.~8, pp. 2004 -- 2031, August 2007.

\bibitem{Parlett}
B.~N. Parlett, \emph{The symmetric eigenvalue problem}.\hskip 1em plus 0.5em
  minus 0.4em\relax NJ, USA: Prentice-Hall, Inc. Upper Saddle River, 1998.

\bibitem{Ben2001}
M.~Bengtsson and B.~Ottersten, \emph{Optimal and Suboptimal Transmit
  Beamforming}.\hskip 1em plus 0.5em minus 0.4em\relax CRC Press, 2001.

\bibitem{Cha2009}
V.~Chandrasekhar, J.~G. Andrews, T.~Muharemovic, Z.~Chen, and A.~Gatherer,
  ``Power control in two-tier femtocell networks,'' \emph{{IEEE} Trans.
  Wireless Commun.}, vol.~8, no.~8, pp. 4316--4328, 2009.

\bibitem{Ha-VuTran2017}
H.-V. Tran, G.~Kaddoum, H.~Tran, and E.-K. Hong, ``Downlink power optimization
  for heterogeneous networks with time reversal-based transmission under
  backhaul limitation,'' \emph{IEEE Access}, vol.~5, pp. 755 -- 770, January
  2017.

\bibitem{Bambos2000}
N.~Bambos, S.~C. Chen, and G.~J. Pottie, ``Channel access algorithms with
  active link protection for wireless communication networks with power
  control,'' \emph{{IEEE/ACM} Trans. Netw.}, vol.~5, no.~8, pp. 583--597,
  October 2000.

\bibitem{Rappaport94}
T.~Rappaport and S.~Sandhu, ``Radio-wave propagation for emerging wireless
  personal-communication systems,'' \emph{{IEEE} Antennas Propag. Mag.},
  vol.~36, no.~5, pp. 14--24, October 1994.

\bibitem{Searle}
S.~R. Searle, \emph{Linear Models}.\hskip 1em plus 0.5em minus 0.4em\relax New
  York: Wiley, 1971.

\end{thebibliography}

\end{document}